\documentstyle[12pt,aasms4]{article}
\begin{document}
\newcommand {\E}   {${\it E}_0$ }
\newcommand {\ie} {{\it i.e.,} }
\newcommand {\eg} {{\it e.g.} }
\newcommand {\etal} {{\it et al.} }

\newcommand{\der}[2]  { \frac{{\rm d}#1}{{\rm d}#2} }
\newcommand{\derp}[2] { \frac{\partial #1}{\partial #2} }
\newcommand{\dif}     {{\rm d}}
\newcommand{\difp}    {\partial}
     
\def\reff{\noindent\hangindent=1pc \hangafter=1}
% Page sizes and formats
%\magnification\magstep1
%
\def\a4{\hsize 15.cm \vsize 23.cm}
\def\skipline{\vskip 10.1pt}
\def\double{\baselineskip 18pt \lineskip 10pt}     % Double spacing for drafts
\def\newpage{\vfill\eject}
\def\under{}
\font\chiquito=cmr9
%
% SYMBOLS
%
\def\Diaz{\hbox{D\'\i az}}
\def\neq{\mathrel{\not=}}
\def\twiddles{\hbox{$sim $}}
\def\varomega{\varpi}
%
% UNITS
%
\def\ang{\thinspace\hbox{\AA}}
\def\km{\thinspace\hbox{km}}
\def\Mpc{\thinspace\hbox{Mpc}}
\def\kpc{\thinspace\hbox{kpc}}
\def\whz{\thinspace\hbox{WHz}}
\def\kms{\thinspace\hbox{$\hbox{km}\thinspace\hbox{s}^{-1}$}}
\def\kmsec{\thinspace\hbox{$\hbox{km}\thinspace\hbox{s}^{-1}$}}
\def\kmsecmeg{\thinspace\kmsec\Mpc$^{-1}$}
\def\erg{\thinspace\hbox{$\hbox{erg}$}}
\def\ergs{\thinspace\hbox{$\hbox{erg}\thinspace\hbox{s}^{-1}$}}
\def\ergsec{\thinspace\hbox{$\hbox{erg}\thinspace\hbox{s}^{-1}$}}
\def\ergsqcmsec{\thinspace\hbox{erg}\sqcm\sec$^{-1}$}
\def\sqcm{\thinspace\hbox{$\hbox{cm}^{2}$}}
\def\cm3{\thinspace\hbox{$\hbox{cm}^{3}$}}
\def\cm2{\thinspace\hbox{$\hbox{cm}^{2}$}}
\def\persqcm{\thinspace\hbox{$\hbox{cm}^{-2}$}}
\def\percc{\thinspace\hbox{$\hbox{cm}^{-3}$}}
\def\percucm{\thinspace\hbox{$\hbox{cm}^{-3}$}}
\def\kev{\thinspace\hbox{keV}}
\def\sec{\thinspace\hbox{s}}
\def\ha{\hbox{$\hbox{H}_\alpha$}}
\def\hb{\hbox{$\hbox{H}_\beta$}}
\def\hg{\hbox{$\hbox{H}_\gamma$}}
\def\hd{\hbox{$\hbox{H}_\delta$}}
\def\lya{\hbox{$\hbox{Ly}_\alpha$}}
\def\kelvin{\thinspace\hbox{K}}
\def\dyncm2{\thinspace\hbox{$\hbox{dyn}\thinspace\hbox{cm}^{-2}$}}
\def\deg{\hbox{$^\circ$}}
\def\degk{\hbox{$^{\circ}K$}}
\def\rstar{\thinspace\hbox{$\hbox{R}_*$}}
\def\vstar{\thinspace\hbox{$\hbox{V}_*$}}
\def\Zsun{\thinspace\hbox{$\hbox{Z}_{\odot}$}}
\def\msun{\thinspace\hbox{$\hbox{M}_{\odot}$}}
\def\rsun{\thinspace\hbox{$\hbox{R}_{\odot}$}}
\def\lsun{\thinspace\hbox{$\hbox{L}_{\odot}$}}
\def\snr{\thinspace\hbox{$\nu_{SN}$}}
\def\gpar{\hbox{$g_{\parallel}$}}
\def\Ca{$\lambda\lambda$\thinspace\hbox{8498,8542,8662}~\AA}
\def\C2{$\lambda\lambda$\thinspace\hbox{8542,8662}~\AA}
\def\Mg{$\lambda$\thinspace\hbox{8807}~\AA}
\def\Fe{\hbox{$(Fe/H)$}}
\def\O{\hbox{$(O/H)$}}
\def\lg{\hbox{$log\thinspace\hbox{g}$}}
\def\lT{\hbox{$log\thinspace\hbox{T}_{eff}$}}
\def\Teff{\hbox{$T_{eff}$}}
\medskip
\def\np{\hfill\vfill\eject\noindent}    % start a new page
\def\nl{\par\noindent}                  % start a new line
\def\t10#1{$\times 10^{#1}$}            % x 10 for scientific notation
\def\10#1{$ 10^{#1}$}                   % 10^ for scientific notation
\def\sol{$_\odot$ }                     % solar mass symbol
\def\Kms{km s$^{-1}$ }                  % velocity units
\def\pcm3{cm$^{-3}$ }                   % number density units
\def\cm2{cm$^{-2}$ }                    % number density units
\def\pyr{yr$^{-1}$ }                    % velocity units
\def\mdot{$\dot {M}_{wind}$ }            % dot M wind
\def\dst{$\dot {N}_{*}$ }               % dot N star
\def\ro{$\rho$}                         % rho
\def\sigg{$\sigma_{gas}$ }               % sigma
\def\sigs{$\sigma_{*}$ }                 % sigma
\def\sig{$\sigma$ }                     % sigma
\def\eps{$\epsilon$}                    % epsilon
\def\reg{$_{region}$}                   % region
\def\alph{$_\alpha$ }
\def\p3{$\pi$}                          % alpha index
\def\st{$_{*}$ }
\def\vst{$v_{*}$}
\def\b{$_{bow shock}$ }
\def\bow{$_{bow}$ }
\def\L{{\it L$_*$} }
\def\sn{$_{sn}$ }                       % sn index
\def\sg{$_{sg}$ }                       % sn index
\def\th{$_{th}$ }                       % thermalization index
\def\kin{$_{k}$ }                       % thermalization index
\def\wnd{$_{wnd}$ }
\def\clo{$_{cloud}$ }
\def\cond{$_{cond}$ }
\def\c{$_{c}$ }
\def\f{$_{f}$ }
\def\cl{$_{cluster}$ }                  % cluster index
\def\vwnd{$v_{wnd}$ }                   % Wind index
\def\gas{$_{gas}$ }
\def\shell{$_{shell}$ }                 % shell index
\def\ejecta{$_{ej}$ }                   % ejecta index
\def\0{$_0$ }                           % 0 initial index
\def\etal{{\it et al.} }                % et al. italicized
\def\si{$\sim$ }                        % similar to
\def\ie{{\it i.e.} }                    % i.e. italicized
\def\cf{{\it c.f.} }                    % c.f. italicized
\def\eg{{\it e.g.} }                    % e.g. italicized
\def\GTT{{\ Tenorio-Tagle} }            % short for my name.
\def\MNR{{\ R\'o\.zyczka} }             % short for Michals name.
\def\CMT{{\ Mu\~noz-Tu\~non} }
% Define standard references
%
%
\def\ApJ#1{{\it Astrophys}. {\it J}. {\bf #1}}
\def\ApJL#1{{\it Astrophys}. {\it J}. {\it Lett.} {\bf #1}}
\def\A+A#1{{\it Astron}. {\it Astrophys}. {\bf #1}}
\def\MNRAS#1{{\it Mon}. {\it Not}. {\it Roy}. {\it Astron}.
             {\it Soc}. {\bf #1}}

%
% REFERENCES
%
\def\reference#1#2#3#4#5{\nine\par\noindent\hangindent 3em
              #1, #2. {\ninesl #3\/}, {\nineb #4,}
\thinspace\hbox{#5}}
\def\refnum#1#2#3#4{\nine\noindent #1, #2. {\ninesl #3\/}, {\nineb #4}}
\def\refpress#1#2#3{\nine\par\noindent\hangindent 3em
             #1, #2. {\ninesl #3\/}, in press}
\def\refsub#1#2#3{\nine\par\noindent\hangindent 3em
           #1, #2. {\ninesl #3\/}, submitted}
\def\refacc#1#2#3{\nine\par\noindent\hangindent 3em
           #1, #2. {\ninesl #3\/}, accepted}
\def\refbook#1#2#3{\nine\par\noindent\hangindent 3em
            #1, #2. {\ninesl #3\/}}
\def\etal   {{\sl et\nobreak\ al.\ }}
\def\etalp   {{\tenib et\nobreak\ al.\ }}
\def\etalr   {{\ninesl et\nobreak\ al.\ }}
\def\aanda  {Astr.\ Astro\-phys.\nobreak\ }
\def\aandas {Astr.\ Astro\-phys.\nobreak\ Suppl.\nobreak\ }
\def\aj     {Astron.\nobreak\ J.\nobreak\ }
\def\annrev {Ann.\ Rev.\ Astr.\ Astro\-phys.\nobreak\ }
\def\acta   {Acta Astron.\nobreak\ }
\def\apj    {Astro\-phys.\nobreak\ J.\nobreak\ }
\def\apjs   {Astro\-phys.\nobreak\ J.\ Suppl.\nobreak\ }
\def\apjl   {Astro\-phys.\nobreak\ J.\ Lett.\nobreak\ }
\def\apspsc {Astro\-phys.\nobreak\ Sp.\nobreak\ Sc.\nobreak\ }
\def\apjsupp{\apjs}
\def\aplett {Astro\-phys.\nobreak\ Lett.}
\def\commap {Comments\ Astrophys.\nobreak\ }
\def\ibvs   {Inf.\ Bull.\ var.\ Stars}
\def\mn     {Mon.\ Not.\ R.\ astr.\nobreak\ Soc.\nobreak\ }
\def\pasp   {Publ.\ astr.\ Soc.\ Pacif.\nobreak\ }
\def\pasj   {Publ.\ astr.\ Soc.\ Japan\nobreak\ }
\def\sovast {Soviet astr.}
\def\jump   {\par\vskip 0.2cm\noindent\hangindent 3em}
%
% Registers
%
\newcount\notenumber
\def\clearnotenumber{\notenumber=0}
\def\note{\advance\notenumber by 1
  \footnote{$~{\the\notenumber}$}}
\newcount\natrefreg
\def\clearnatref{\natrefreg=0}
\def\natref{\advance\natrefreg by 1 $~{\the\natrefreg}$}
\newcount\secreg
\def\clearsecreg{\secreg=0}
\newcount\subsecreg
\def\clearsubsecreg{\subsecreg=0}
\def\newsection#1{\skipline\skipline\par\smallbreak\noindent\advance\secreg by1
 \the\secreg . {\bf #1}\nobreak\par\clearsubsecreg}
\newcount\tablereg
\def\cleartable{\tablereg=0}
\def\nextable{\advance\tablereg by1\table}
\newcount\figreg
\def\clearfig{\figreg=0}
\def\fig{Fig.~\the\figreg}
\def\nextfig{\advance\figreg by1\fig}
%
% Abbreviations
%
\def\rsa{{\ninesl Revised Shap\-ley-Ames Catalog of Bright Gal\-axies\/} Sandage
 \&\ Tammann (1981)}
\def\RC2{{\ninesl Second Reference Catalogue of Bright Gal\-axies\/},
 de~Vaucouleurs, de~Vaucouleurs and Corwin (1976)}
\def\rc2{\RC2}
\def\Sin{\hbox{Sin}}
\def\Cos{\hbox{Cos}}
%
% Other macros
%
\def\deriv#1#2{\hbox{${{\displaystyle\hbox{d}#1}\over
{\displaystyle\hbox{d}#2}}$}}
\def\sderiv#1#2{\hbox{${{\displaystyle\hbox{d}^2#1}\over
{\displaystyle\hbox{d}#2^2}}$}}
\def\pderiv#1#2{\hbox{${{\displaystyle\partial#1}\over
{\displaystyle\partial#2}}$}}
\def\spderiv#1#2{\hbox{${{\displaystyle\partial^2#1}\over
{\displaystyle\partial#2^2}}$}}
\def\x10#1{\hbox{$\times\hbox{10}^{#1}$}}
\def\eex#1{\hbox{$\hbox{10}^{#1}$}}
\def\movements#1{\halign{\quad\it##\hfil&&\qquad\it##\hfil\cr#1\crcr}}
%
% Reset default
%
\def\integ{\int\limits}
% for ApJ format remove comment in next setence
%\double

%MMMMMMMMMMMMMMMMMMMMMMM  MACROS & DEFINTIONS MMMMMMMMMMMMMMMMM
\def\ni{\noindent}                                       %No indent%
\def\ls{\vskip 12.045pt}                            %One Line space%
\def\et{et\thinspace al.\ }                                 %et al.%

\def\gapprox{$_>\atop{^\sim}$}     %Greater than over approximately%
\def\lapprox{$_<\atop{^\sim}$}        %Less than over approximately%

\def\kms{km\thinspace s$^{-1}$}                        %km per sec1%
\def\ergsec{erg\thinspace s$^{-1}$}                  %ergs per sec%
\def\msun{M$_{\odot}$}                                  %Solar mass%
\def\cms{{\rm cm\thinspace s}$^{-1}$}                   %cm per sec%
\def\percc{cm$^{-3}$}                         %per cubic centimeter%
\def\peryr{yr$^{-1}$}                                     %per year%
\def\perday{day$^{-1}$}                                    %per day%
\def\persec{s$^{-1}$}                                   %per second%

\def\Ha{$H\alpha$}                                         %H alpha%
\def\Hb{$H\beta$}                                           %H beta%

\def\beq{\begin{equation}}                          %begin equation%
\def\eeq{\end{equation}}                              %end equation%
\def\beqa{\begin{eqnarray}}                         %begin eqnarray%
\def\eeqa{\end{eqnarray}}                             %end eqnarray%
\def\beqan{\begin{eqnarray*}}                      %begin eqnarray*%
\def\eeqan{\end{eqnarray*}}                          %end eqnarray*%

\newcommand{\ET}[1]{\times 10^{#1}}             %times power of ten%

\newcommand{\ov}[1]{\overline{#1}}                        %overline%
\newcommand{\zb}[1]{\left[ {#1} \right]}           %square brackets%
\newcommand{\zk}[1]{\left\{ {#1} \right\}}                    %keys%
\newcommand{\zp}[1]{\left( {#1} \right)}               %parentheses%

\def\ojo{\fbox{\bf !`ojo!}}        %OJO! Something Needs Correction%
%MMMMMMMMMMMMMMMMMMMMMMMMMMMMMMMMMMMMMMMMMMMMMMMMMMMMMMMMMMMMMMMMMMM

\title
{\bf On the Fate of Processed Matter in Dwarf Galaxies}

\def\runninghead
{ SILICH and TENORIO-TAGLE:\quad Processed Matter in  Dwarf Galaxies}

% VERSION 21, Cambridge, May 26 1997
%
\author{ Sergey A.
Silich\altaffilmark{1,5} \& Guillermo Tenorio-Tagle\altaffilmark{2,3,4}}

\altaffiltext{1}{Main Astronomical Observatory National Academy 
of Sciences of Ukraine, 252650, Kiev, Golosiiv, Ukraine. 
(silich@mao.kiev.ua)}

\altaffiltext{2}{INAOE. Apartado Postal 51, 72000 Puebla, M\'exico.}

\altaffiltext{3}{
Institute of Astronomy, University of Cambridge, Madingley Road, Cambridge 
CB3 0HA, UK}

\altaffiltext{4}{Royal Greenwich Observatory, Madingley Road, Cambridge 
CB3 OEZ, UK} 

\altaffiltext{5}{Invited participant at the ``Starburst Workshop 96'' 
under the 
``Guillermo Haro International Program for Advanced Astrophysical Research''.}
 
\date{
Received $\underline{\ \ \ \ \ \ \ \ \ \ \ \ \ \ \ \ \ \ \ \ \ \ }$ \ \ 
Accepted $\underline{\ \ \ \ \ \ \ \ \ \ \ \ \ \ \ \ \ \ \ \ \ \ }$ \\
} 
%\maketitle
%\vspace{-1.2cm} % include according to taste.
%----------------------------------------------------------------------

\begin{abstract}

\noindent Two dimensional calculations of the evolution of remnants 
generated by the strong mechanical energy deposited by stellar clusters in 
dwarf galaxies (M \si $10^9 - 10^{10}$ \msun) are presented. The evolution 
is followed for times longer than both the blowout time and the presumed 
span of energy injection generated by a coeval massive stellar cluster. The 
remnants are shown to end up wrapping around the central region of the host 
galaxy, while growing to kpc-scale dimensions. Properties of the remnants 
such as luminosity, size, swept up mass, and expansion speed are given as 
a function of time for all calculated cases. 

The final fate of the swept-up galactic gas and of the matter processed by 
the central starburst is shown to be highly-dependent on the properties of 
the low density galactic halo. Superbubbles powered by star clusters, with 
properties similar to those inferred from the observations, slow down  in 
the presence of an extended halo to expansion speeds smaller than the host
galaxy escape velocity. Values of the critical luminosity required for the 
superbubbles to reach the edge of the galaxies with a speed comparable to 
the escape speed are derived analytically and numerically. The critical 
luminosities are larger than those in the detected sources and thus, the 
superbubbles in amorphous dwarf galaxies must have already undergone blowout 
and are presently evolving into  an extended low density halo. This will 
inhibit the loss of the swept-up and processed matter from the galaxy.  

\end{abstract}

\section {Introduction} 
Massive stellar clusters violently eject their processed matter 
through   supernova (SN) explosions and
strong stellar winds, leading to the formation of large-scale
remnants commonly known as superbubbles. These,  when generated  
in dwarf galaxies,  are thought to acquire a dimension that rapidly exceeds,
at least along the galaxy minor axis, the scale height of the mass
distribution. When this occurs, the remnants are believed to 
blow out of their host  galaxy \ie as the shell of swept up  
interstellar matter (ISM) accelerates outwards and  fragments via 
Rayleigh-Taylor (R-T) instabilities, it  allows the processed matter  
to freely stream as a wind into the intergalactic medium (IGM). 
Furthermore, the most energetic events are thought to cause not only the 
loss of their newly processed metals but to  be able to strip dwarf galaxies 
of their entire ISM (see De Young \& Heckman 1994 and references therein).

Here however, we show that despite the fact that remnants easely 
manage to blowout or burst out of the disk matter distribution, 
under a wide variety of
circumstances the loss of matter could be  inhibited by the hydrodynamic
response of the rapidly streaming wind gas to the conditions found in the
galaxy halo. 
  
Analytical and numerical hydrodynamic studies of the
remnants caused by a strong stellar energy input in a plane-stratified
atmosphere have analyzed the blowout phenomenon (Kompaneets 1960;
\GTT \& Bodenheimer 1988; Mac Low et al. 1989; Igumentshehev, Tutukov
\& Shustov 1990; Tenorio-Tagle, \MNR \& Bodenheimer 1990). For a recent 
account see Koo \& McKee (1992) who derived the characteristic wind 
luminosity ($L_b$) that determines whether a bubble remnant is able to 
blowout of the disk configuration. $L_b$ is the luminosity required for the 
expanding shell of swept up matter to
have a sonic velocity when it reaches a disk scale height. 
$L_b = 2.4 \times 10^{36} P_4 H_{100}^2 a_{s,10}$  erg s$^{-1}$,
where $P_4 = P_0/(10^4k)$ \percc K, $H_{100}$ is the disk scale height in
units of 100 pc and $a_{s,10}$ is the disk sound speed in units of 
10 \kms.  Koo \& McKee (1992) estimated also that if 
the cluster luminosity ($L_*$) exceeds $3L_b$ the bubble would  expand
supersonically when reaching a scale height and it will begin to accelerate 
to blowout of the
disk into the halo of the galaxy. Once the bubble blows out, 
there are two possible end solutions depending on the wind luminosity.
For $L_* >> L_b$ blowout would cause the venting of the hot bubble
interior out of the planar atmosphere.  Lower wind 
luminosities ($3 \le L_*/L_b \le 10$) result in the formation of stable
hydrodynamical jets as the processed  matter is re-shocked and accelerated to
supersonic speeds through a nozzle generated by the 
shocked disk matter.  In this particular case there is no doubt the
processed matter is due to escape the galactic system.  The more
energetic events ( $L_* >> L_b$ ) destroy upon expansion the possibility of a 
nozzle formation and the hot gas in such cases acquires a maximum speed
similar to its own velocity of sound, as it streams through the fragmented 
shell into the halo.  
In dwarf galaxies, these remnants have recently been
investigated by several groups. In particular, 
amorphous dwarf galaxies 
have been selected as ideal laboratories to study the impact of 
starbursts on the ISM, and to investigate the possible
ejection and loss of their ISM and  processed matter into the IGM. 
These galaxies are ideal laboratories because they have 
an ISM mass in the range    10$^7$ - 10$^{9}$
M\sol and a concentrated central starburst, as opposed to Magellanic 
Irregulars where the star formation seems to extend over the whole body
of the galaxies (Gallagher \& Hunter 1987; 
Marlowe, Heckman, Wise \& Schommer 1995
and references therein). Theoretically these issues have been
address by  De Young \&
Gallagher (1990) who estimated that about 2/3 of the processed matter
ejected by supernova will escape the galaxy. This conclusion was reached 
from an analysis of  2-D numerical calculations in a 1 kpc$^2$ grid, after 
comparing the processed matter expansion speed 
with the escape velocity of their selected galaxy.
In a further study, 
De Young \& Heckman (1994), who considered dwarf galaxies with masses in
the range $10^7 - 10^9$ M\sol and physical sizes of 1 - 3 kpc, found
that the degree of flatness of the assumed density distribution is an
important parameter. This is because after shock acceleration and 
through the blowout phenomena, the
energy of the starburst tends to escape along the galaxy minor axis
imparting little momentum to the gas in the outer portions of the disk
matter distributions. Thick disks however, were found to be completely
disrupted and accelerated to speeds larger than the  escape 
velocity of galaxies. Note however, that the approach of De Young \& Heckman 
does not include the contribution of dark matter to the gravitational 
potential of the dwarf galaxy, nor the existense of an extended halo component.
Note also that the range of energies used 
spans to up to 10$^{44}$ erg s$^{-1}$, although the maximum assumed energy 
was never allowed to surpass the number of SN expected if the whole 
galaxy ISM ($M_{ISM}$) was turned into stars following a Salpeter IMF \ie
the maximum used number of supernovae  N$_{SN}$ = M$_{ISM}$/100.  It is 
important to note that the best studied dwarf galaxies have an estimated 
mechanical energy deposition in the range of 10$^{40}$ - 10$^{41}$ 
erg s$^{-1}$ (Marlowe \etal 1995). This implies a SN rate of 
$3 \times (10^{-4} - 10^{-3})$ per year and a total expected energy 
injection of $1.3 \times (10^{55} - 10^{56})$ erg if the energy input rate 
is assumed to last for 4 $\times$ 10$^7$ yr, time after which all massive 
stars from a coeval starburst would have exploded as SN.   

Unfortunately, there is little information about the gaseous
structure of dwarf galaxies. However, in all 
well studied cases the size of the region photoionized by the cluster of 
massive stars, exceeds the size of the superbubbles present in these 
galaxies (see Marlowe \etal 1995; their tables 4 and 5). Furthermore, the 
few that have been looked at in HI (see for example Hoffman \etal 1993; 
Hunter, Van Worden and Gallagher, 1994; Meurer 1994) display large envelopes,
several times larger than the
optical size of the galaxies. One can obviously argue that it may just be a
projection effect, however,  the galaxies under consideration are
of the "amorphous" type  and thus it is rather unlikely that all
observed matter is confined to a narrow disk. 
It thus seems perhaps  more likely that the superbubbles observed in these
galaxies  (Marlowe \etal, 1995; Martin 1996) which 
present a continuous horseshoe shape attached to the central star forming 
region and which have  dimensions \si 1 kpc and expansion velocities \si 50 \kms are 
presently evolving into the extended halos of their host galaxies. 

Here we present two dimensional calculations of multi-supernova
remnants evolving in dwarf galaxies (see sections 2 and 3). We find a good
agreement with the above mentioned studies although our 
conclusion (see sections 4 and 5) is that the final fate of the swept-up 
and processed matter is highly-dependent on the properties of the previously
ignored galactic gaseous halos.

%and  thus, for the  conditions derived from the 
%observations, galaxies with a gas mass around 10$^9$ M\sol, or larger, do 
%not loose their ISM, nor their newly processed metals. 

\section{Dwarf Galaxies}

\subsection{The ambient gas density distribution}

Massive galaxies ($M_{Gal}$ $\geq$ 10$^{10}$ M\sol), such as the Milky Way,
are capable of retaining a hot ($T_{Gal}$ \si a few 10$^6$K) 
and extended massive halo completely bound to their  
gravitational potential. This is frequently re-heated by 
a collection of  superbubbles, caused by aging OB associations, 
that blowout of the plane HI disk density stratification.    
Less massive galaxies, on the other hand, are unable to keep such massive and
hot halos. Instead, the smaller the galaxy total mass, the cooler and less 
extended their possible resultant halos become. 
Thus these are usually idealized as a low density turbulent medium 
(or clouds) that extends up to a maximum radius R$_G$ at which its 
characteristic velocity dispersion exceeds the galaxy escape speed.
 As stated by Suchkov 
\etal (1994, hereafter SBHL), "in a real galaxy, much of the hydrostatic 
support of the interstellar matter is provided by the random motions of 
interstellar clouds". 

Following Li \& Ikeuchi (1992), Tomisaka \& Ikeuchi (1988, hereafter TI), 
Tomisaka \& Bregman (1993, hereafter TB), and SBHL,  we model our dwarf 
galaxies with three different isothermal components. 
%--------------------------------------------------------------------
\begin{equation}
      \label{eq.1}
\rho_{g} = \rho_{NM} +  \rho_{IM} + <\rho_{halo}>
\end{equation}
%--------------------------------------------------------------------
where $\rho_{NM}$, $\rho_{IM}$ and $<\rho_{halo}>$ are the gas densities
of the neutral and ionized components and the mean density of the 
extended turbulent halo material, respectively. 
The model corresponds to the bound-cool type halo in the notation of Li \&
Ikeuchi (1992). The total gas pressure is then given by
%--------------------------------------------------------------------
\begin{equation}
      \label{eq.2}
P_{ext} = \frac{1}{\gamma} \rho_{NM} C_{NM}^2 +
\frac{1}{\gamma} \rho_{IM} C_{IM}^2 +
\frac{1}{3} <\rho_{halo}> C_{halo}^2,
\end{equation}
%-------------------------------------------------------------------- 
where $C_{NM}$, $C_{IM}$ and $C_{halo}$ are the sound velocities of
the neutral and ionized gas and the velocity dispersion of the 
extended halo component. $\gamma$ is the ratio of specific heats.

Let $R_B$ be the maximum extent of the stars and dark matter, and $V_B$, the
escape velocity at this surface. Following TI, and TB we assume that 
a fraction $(1-e^2)$ of the radial component of gravity is balanced by 
pressure gradients. Then the initial gas density distribution follows
from the relation: 
%--------------------------------------------------------------------
\begin{equation}
      \label{eq.3}
\frac{\rho_{g}}{\rho_0} = \alpha_{NM}
\exp{\left[\frac{\gamma}{2}\left(\frac{V_B}{C_{NM}}\right)^2 \chi
\right]} + \alpha_{IM} \exp{\left[
\frac{\gamma}{2}\left(\frac{V_B}{C_{IM}}\right)^2 \chi \right]} + 
\alpha_{halo} \exp{\left[\frac{3}{2}\left(\frac{V_B}{C_{halo}}\right)^2
\chi \right]} ,
%[F - e^2 F(0,r) - (1-e^2) F(0,0)]}
\end{equation}
%--------------------------------------------------------------------
where $\rho_0$ is the total gas density at the galactic center,
$\alpha_{NM}=\rho_{NM,0}/\rho_0$, $\alpha_{IM}=\rho_{IM,0}/\rho_0$ and
$\alpha_{halo}=<\rho_{halo,0}>/\rho_0$ are the density ratios of the 
various components to the total density at the galactic
center. The function $\chi$ equals
%--------------------------------------------------------------------
\begin{equation}
      \label{eq.4}
\chi = F(\omega) - e^2 F(r) - (1-e^2) F(0),
\end{equation}
%-------------------------------------------------------------------- 
where $r=\sqrt{x^2 + y^2}$ and $\omega = \sqrt{x^2 + y^2 + z^2}$ are the
cylindrical and spherical radii, and the function $F$ is defined by 
equation (\ref{eq.9}). The rotation velocity of the ISM component comes 
from the relation:
%--------------------------------------------------------------------
\begin{equation}
      \label{eq.4a}
\frac{V^2}{r} = e^2 (g_x^2 + g_y^2)^{1/2},
\end{equation}
%-------------------------------------------------------------------- 
where $g_x$ and $g_y$ are the x and y components of the gravity field.

We include a transition from a rotating disk component to a non-rotating
spherical halo and radial gradients in the disk density distribution by
assuming 
%--------------------------------------------------------------------
\begin{equation}
      \label{eq.5}
e = e_0 / \exp\left[\left(\frac{z}{H_z}\right)^2 +
\left(\frac{r}{H_r}\right)^2 \right] ,
\end{equation}
%-------------------------------------------------------------------- 
where $H_r$ and $H_z$ are the characteristic scale heights
along and perpendicular to the plane of the galaxy. Then 
(\ref{eq.3}) is not self-consistent if a hydrostatic gaseous disk is
assumed, but it is asymptotically  correct for a hydrostatic gaseous
halo (TI, SBHL). Typical  dwarf galaxy  density distributions assuming a 
total mass M$_{B}$ = 10$^{10}$ M\sol (models A and B) and 
10$^9$ M\sol (Models C) and a 10$\%$ gaseous component 
are shown in the Figure 1 (see also Table 1). Note that the last 
contour drawed represents, in each case, the gaseous 
edge of the galaxy ($R_G$). 
\ie material further out with a velocity dispersion 
$C_{halo}$ would exceed the escape velocity and thus it would not be bound
to the system. Figure 2 gives a global impression of how the galactic ISM is
distributed for different values of the assumed velocity dispersion of the 
extended component (also used to identify the models; see Table 1) 
and thus it shows how for larger values of $C_{halo}$,
the smaller, or more compact that the bound gaseous component results.
 The value of P$_{ext}$($R_G$)/k varies in our models within a range
    0.8 cm$^{-3}$ K (Model B60) - 5.4$\times$ 10$^4$ cm$^{-3}$ K
(Model A120). This pressure should balance the  extragalactic gas 
pressure and in fact covers the range often used in the  literature 
(eg. Babul \& Rees 1992).

Several calculations were also made for different ratios of $M_{ISM}/M_{B}$, 
as well as for different dimensions  of the dark matter distribution 
(see section 5).

\subsection{The gravitational field}

We have assumed that the main contribution to the gravitational field 
results from spherically symmetric stellar and dark matter components, 
disregarding the contribution of a stellar disk component, 
and the self-gravity of the gaseous matter.
A King model was assumed for the background stars and a distribution inversely 
proportional to the square of the radius has been adopted for a dark matter 
halo:
%-------------------------------------------------------------------- 
\begin{equation}
\label{eq.6} \rho_{tot}(\omega) = \frac{\rho_1} {\left[1 +
\left(\frac{\omega}{R_1} \right)^2 \right]^{3/2}} +
\frac{\rho_2} {1 + \left(\frac{\omega}{R_2} \right)^2},
\end{equation}
%-------------------------------------------------------------------- 
where $R_1$ and $R_2$, $\rho_1$ and $\rho_2$ are characteristic
scales and densities of the star and dark matter components, respectively.

The total mass of these components may be expressed as 
%-------------------------------------------------------------------- 
\begin{eqnarray}
\label{eq.7} 
& & \hspace{-0.5cm}
M_B = 4 \pi \int_0^{R_B} \rho_{tot}(\omega)\omega^2{\rm
d}\omega =
\nonumber
\\[0.2cm] & & \hspace{-0.5cm}
M_1 \left[\ln{(x_B + \sqrt{1+x_B^2})} - \frac{x_B}{\sqrt{1+x_B^2}}
\right] +
M_2 \left[y_B - \arctan{(y_B)}\right],
\end{eqnarray} 
%-------------------------------------------------------------------- 
where $x_B = R_B/R_1$, $M_1 = 4 \pi \rho_1 R_1^3$, $y_B = R_B/R_2$,
$M_2 = 4 \pi \rho_2 R_2^3$. We fix a constant at the gravity potential
by assuming that it becomes Newtonian at the galactic boundary $R_B$.
Then the gravity potential corresponding to the
adopted total density distribution is expressed as follows:
%--------------------n------------------------------------------------ 
\begin{equation}
\label{eq.8} 
\Phi(r,z) = -\frac{V_B^2}{2} F(\omega),
\end{equation}
%-------------------------------------------------------------------- 
where function $F(\omega)$, is defined by the expression (see also SBHL)
%-------------------------------------------------------------------- 
\begin{eqnarray}
\label{eq.9} 
& & \hspace{-0.5cm}
F(\omega) = 1 + \frac{R_B}{R_2} \frac{M_2}{M_{tot}}\left[
\frac{1}{2}\left(\ln{(1+y_B^2)} - \ln{(1+y_{\omega}^2)}\right) +
\frac{\arctan{(y_B)}}{y_B} - \frac{\arctan{(y_{\omega})}}{y_{\omega}}\right] +
\nonumber
\\[0.2cm] & & \hspace{-0.5cm}
 \frac{R_B}{R_1} \frac{M_1}{M_{tot}}\left[
\frac{\ln{(x_{\omega}+\sqrt{1+x_{\omega}^2}})}{x_{\omega}} -
\frac{\ln{(x_B+\sqrt{1+x_B^2}}}{x_B} 
\right], \quad \omega \le R_B,
\\[0.2cm] & & \hspace{-0.5cm}
F(\omega) = \frac{R_B}{\omega}, \quad \omega > R_B, \nonumber
\end{eqnarray} 
%-------------------------------------------------------------------- 
where $x_{\omega} = \omega/R_1$, $y_{\omega} = \omega/R_2$.
The value of $V_B$ is given by
%-------------------------------------------------------------------- 
\begin{equation}
\label{eq.10} 
V_B = \sqrt{\frac{2GM_B}{R_B}},
\end{equation}
%-------------------------------------------------------------------- 
whereas the escape velocity at the current position $(r,z)$ is
%-------------------------------------------------------------------- 
\begin{equation}
\label{eq.11} 
V_{esc} = V_B\sqrt{F(\omega)}.
\end{equation}
%-------------------------------------------------------------------- 

The $x,y,z$ components of gravity for $\omega \le R_B$ 
derived from (\ref{eq.8}) and (\ref{eq.9}) are
%-------------------------------------------------------------------- 
\begin{eqnarray}
\label{eq.12} & & \hspace{-0.5cm} 
g_{x}=- \frac{G M_1}{\omega^2}
\frac{x}{\omega}
\left[\ln{(x_{\omega}+\sqrt{1+x_{\omega}^2})} -
\frac{x_{\omega}}{\sqrt{1+x_{\omega}^2}} + 
\frac{M_2}{M_1}(y_{\omega}-\arctan{y_{\omega}})\right], 
\\[0.2cm] & & \hspace{-0.5cm} 
g_{y}=- \frac{G M_1}{\omega^2}
\frac{y}{\omega} \left[\ln{(x_{\omega}+\sqrt{1+x_{\omega}^2})} -
\frac{x_{\omega}}{\sqrt{1+x_{\omega}^2}} + 
\frac{M_2}{M_1}(y_{\omega}-\arctan{y_{\omega}})\right], 
\\[0.2cm] & & \hspace{-0.5cm} 
g_{z}=- \frac{G M_1}{\omega^2}
\frac{z}{\omega} \left[\ln{(x_{\omega}+\sqrt{1+x_{\omega}^2})} -
\frac{x_{\omega}}{\sqrt{1+x_{\omega}^2}} + 
\frac{M_2}{M_1}(y_{\omega}-\arctan{y_{\omega}})\right]. 
\end{eqnarray} 
%-------------------------------------------------------------------- 
For larger distances $\omega > R_B$ they are simply 
%--------------------------------------------------------------------
\begin{eqnarray} \label{eq.13} & & \hspace{-0.5cm} 
g_{x}=- \frac{GM_B}{\omega^2} 
\frac{x}{\omega}, \\[0.2cm] & & \hspace{-0.5cm} g_{y}=- \frac{G
M_B}{\omega^2} 
\frac{y}{\omega}, \\[0.2cm] & & \hspace{-0.5cm} g_{z}=- \frac{G
M_B}{\omega^2} \frac{z}{\omega}.  
\end{eqnarray}
%-------------------------------------------------------------------- 

\section{The Calculations}

Here we present two-dimensional calculations of the evolution of the
remnants generated by multi-supernova explosions from an aging massive
stellar cluster evolving in a low-mass galaxy ($M_{B}$ $\leq 10^{10}$M\sol).
The remnant evolution is calculated by means of the thin
layer approximation (Kompaneets 1960) a method developed in 2D and 3D
by Bisnovatyi-Kogan \& Blinnikov (1982); Mac Low \& McCray (1988); 
Bisnovatyi-Kogan \& Silich (1991); Palou\v{s} (1992); Silich (1992) and
Silich \etal (1996), and now applied to a variety of cases
(Bisnovatyi-Kogan \& Silich 1995).  The evolution is followed up to 60
- 170 Myr, long past the time at which the SN activity from the
evolved - assumed coeval - massive cluster has come to an end. 

\subsection{The energy input} 

Based on the observations of dwarf galaxies (see Marlowe \etal 1995
and references therein) and the synthetic properties of starburst
galaxies derived by Leitherer \& Heckman (1995), 
the energy input rates for the calculations 
were set within the range $L_*$ = 10$^{40}$ - 10$^{41}$ erg s$^{-1}$, 
although in
section 5 much more energetic clusters are considered.  
The assumed  constant energy input rate $L_*$ depends on the total
energy released by massive stars ($E_{burst}$) and the duration of the 
supernova phase ($t_{burst}$) and  can be written as
%--------------------------------------------------------------------
\begin{equation}
      \label{eq.14}
L_{38} = 3.17 \times 10^3 \frac{E_{burst}/10^{56}ergs}{t_{burst}/10^7
year}.
\end{equation}
%--------------------------------------------------------------------
Here the energy supply rate is $L_{38} = L_*$ / $10^{38}$ erg s$^{-1}$. We 
have also assumed that massive stars release an average $m_{SN}=10 M_{\odot}$
during their explosion, and that this is dispersed into the hot 
superbubble interior.
Thus the rate of mass ejecta due to SN explosions, ${\dot
M}_{SN}$, is assumed to be
%--------------------------------------------------------------------
\begin{equation}
      \label{eq.15}
{\dot M}_{SN} = \frac{N_{SN} m_{SN}}{t_{burst}} = \frac{L_* \, m_{SN}}
{E_{SN}}
\end{equation}
%--------------------------------------------------------------------
where $E_{SN}$ is the energy released by each SN explosion
(typically $E_{SN}=10^{51} erg$).
  
\subsection{Initial conditions}

All the calculations start with a small adiabatic spherical bubble of
initial radius $R_e$ at the time 
%--------------------------------------------------------------------
\begin{equation}
      \label{eq.16}
t_0 = \left[\frac{2 \pi (9\gamma + 5)}{75(\gamma-1)}
\frac{\rho_{g}(x_0,y_0,z_0) R_e^5}{L_*} \right]^{1/3}.
\end{equation}
%---------------------------------------------------------------------
The initial expansion velocity, thermal energy, and mass within the bubble are
%--------------------------------------------------------------------
\begin{eqnarray}
      & & \hspace{-0.5cm}
      \label{eq.17}
u_0 = 0.6 \frac{R_e}{t_0},
      \\[0.2cm]
      & & \hspace{-0.5cm}
E_{th}(t_0) = \frac{14}{9\gamma +5} L_*  t_0,
       \\[0.2cm]
      & & \hspace{-0.5cm}
M_{in}(t_0) = {\dot M}_{SN} t_0.
     \end{eqnarray}
%---------------------------------------------------------------------

Table 1 provides the main parameters of the simulated galaxies,
as well as the energetics of their assumed bursts of star formation.
Several parameters were unchanged in the calculations \eg
the metallicity of the galactic gas has been assumed to be $\xi =
Z/Z_{\odot} = 0.3$; the duration of the supernova phase 
$t_{burst}=40\,Myr$; the constant $e_0$ = 0.9 in (eq \ref{eq.5}). 
The parameter $n_{NM,0}=20\,cm^{-3}$, (see (\ref{eq.3}))
the characteristic scale heights for the gas density distribution
have been taken as H$_z = 1$kpc,  H$_r = 0.5 R_B$, whereas the  
characteristic scale height for the King's mass distribution was taken
to be $R_1 = 0.75$ kpc, and the parameter $R_2$ in equation 
(\ref{eq.6}) has been kept as half the truncation radius $R_B$.    
The temperature of the ionized component
was taken as $2 \times 10^4 \, K$ throughout the calculations. 
To satisfy the high pressure condition at the starburst region
we also set the neutral gas temperature equals to $6 \times 10^3 \,
K$. We relaxed  the high pressure constraint for the low mass galaxies 
(models C), for which a  temperature for the neutral gas of 100 $K$ 
was assumed.

 The method of solution and the procedure to calculate  
various remnant properties such as the bolometric and X-ray luminosities  
are given in Appendices A and B. 

\section{A typical evolutionary sequence}

Figure 3 shows the evolutionary sequence of models A100 
(see Table 1)  for which
the initial  ISM density follows the distribution shown in Figure 1b with a 
central density value of 20.2 \percc and a halo velocity dispersion of 100 
\kms. The central starburst was assumed to 
deposit $7.9 \times 10^{40}$ erg s$^{-1}$ during 
the first 40 Myr of evolution. 
As expected, the remnant grows at first elongated in the direction
perpendicular to the disk of the galaxy and blows out into the halo.  Soon
however, as a result of blowout a secondary shell of newly swept up 
halo gas forms. This results from  the stream of the high 
pressure processed matter that bursts out of the superbubble 
interior while causing a symmetrical and larger twin superbubble(see Figure 
3a). 
This immediately begins to wrap around and ends up  surrounding the inner 
densest part of the disk as it grows to large dimensions into the halo. 
Note that more elongated structures  result if steeper
density distributions are assumed (see Figure 3b).
 At later times, the leading shock manages to plough
through the outer less dense sectors of the central disk and merges with
its symmetrical counterpart to form a continuous structure and a 
more spherical single giant superbubble. During the process of shock
merging, a considerable amount of the shell mass
is left behind strangling a dense toroid carved in the initial density 
distribution. All of this matter stops then
participating in the general outward motion and is instead
strongly compressed towards the galaxy plane to remain completely engulfed by 
the almost spherical giant superbubble.   

Figure 4 shows the total amount of matter swept by the remnant as a
function of time. This steadily grows to values of several times
$\sim 10^8$M\sol with
most of the material being collected as the remnant evolves into the
halo. The sharp decrease shown in Figure 4a at about 25 $\times 10^6$ yr
is due to the process of outer shock merging above described. 
Figure 4 also shows the total amount of matter evaporated from 
the outer shell into the remnant interior, and the amount of material
injected by the collection of SN explosions. In all cases, 
the evaporated mass amounts to only  a few percent of the swept up mass
but it constitutes the most significant mass input into the superbubble 
interior; at least an order of magnitude larger than that provided by the
sequential SNe. 

Meanwhile, the hot bubble interior, as well as the shell of swept up
matter, cause the X-ray emission from the nebulae. Figure 4e displays
the total X-ray emission (see Appendix B), and the shell contributions 
to this luminosity. The bolometric bubble luminosity is shown in Figure 4d.
In the conditions of the A100 model the remnant evolves mainly in the
radiative mode and thus the shell contribution to the X-ray emission is
negligible compared to the X-ray emission from the hot superbubble
interior. 

\section {The fate of the processed matter}

The central issue however, is whether or not one can estimate without 
uncertainties the possibility of mass ejection from the galaxy.  For this,
the obvious and clear cut procedure is to compare the fastest 
expansion speed of the remnant (say along the symmetry axis) with the
galaxy escape velocity at different radii, or evolutionary times.  The
maximum speed (u$_{exp}$) of the sector of the remnant evolving into the 
steepest density gradient shows at first a continuous deceleration
(see Figure 5), for $R$'s less than 1 kpc.  Then during blowout into
the galaxy halo,
it suffers a sudden increase towards large speeds  surpasing even the 
galaxy escape velocity values (dashed lines in Figure 5). 
The dynamic halo 
however, makes eventually an impact on the evolving superbubble forcing it 
to slow down again, once a significant
amount of halo material has been incorporated into the remnant 
shell. These trends are shown in Figure 5 where the largest expansion
 remnant speeds (solid lines) for all A and B 
cases (see Table 1, and Figures 1 and  2) 
are compared
with the galaxy escape velocity.
Note that the maximum expansion speed after blowout into the halo,
as well as the slowest deceleration occurs for the 
more spread out distributions of matter or less compact galaxies
considered; which are those with the smallest $C_{halo}$ values.
  All superbubbles attain large expansion speeds, even larger 
than $V_{esc}$, however, they all end up 
with a velocity well below the galaxy escape speed. 

An important point  to notice here is that just as old supernova remnants
are disrupted once their expansion speed drops below the ISM
random speed of motions, once a superbubble expansion 
becomes slower than the assumed velocity dispersion of the host galaxy halo,
it should begin to be disrupted and eventually loose its identity.
Giant, kpc-scale, filaments associated with the host galaxy, may be the only
long lasting trace of the energetics from old stellar generations and 
their disrupted giant remnants. 
Thus important during and after disruption 
is to decide how much further can the hot 
superbubble interior expand. Note that its expansion would lead to 
further cooling and this to a further drop in pressure. The final fate, \ie
whether it will be lost or trapped by the galaxy, can be addressed by comparing
its pressure ($P_{in}$) to the host galaxy outer boundary pressure 
P$_G$ = P$_{ext}$(R$_G$). With this aim in mind, the calculations were 
continued for times that largely exceeded the onset of shell disruption (ie. 
the time when the expansion speed becomes smaller than the random
speed of motions in the halo), and 
which provided us with a handling on the run of P$_{in}$. This, as shown in 
Figure 6, in most cases becomes smaller than P$_G$ with 
the implication that the processed matter 
will remain bound to the system. For the case A60, although $P_{in}$ exceeds 
$P_G$, it does not exceed the local pressure value at the shell stall radius
(dotted line in Figure 6),
and this leads to the contraction of the superbubble and the retention 
of the processed material.   

Table 1 indicates several other calculated cases 
for which we assumed for example a different proportion of dark matter with 
respect to $M_{ISM}$ (D and E cases), or runs without thermal evaporation 
(A*100), or with a lower total mass (C cases). The results are nevertheless
very similar to the above described cases.  

Clearly, the expulsion of matter from a dwarf galaxy will only be 
possible if one could freely increase the energy input rate ($L$). 
The problem however, 
is to know which is the correct or sufficient 
amount for each of the galaxies. A good 
formalism  could be device if one notices from  
Figures 1 and 2 that the extended low density halos are in fact 
rather uniformly spread out and that they constitute a large fraction of the 
galaxies ISM. Thus, given the continuous energy input rate (at least 
for the first 40 Myr) one would expect that the relationships developed for
the evolution of stellar wind bubbles 
in a uniform density medium  (see Weaver \etal 1977) will 
give  reasonable estimates. The functions R(t) and  dR/dt lead to the 
time independent expression   
%--------------------------------------------------------------------
\begin{equation}
      \label{eq.18}
R = 16.6 \left(\frac{L_{38}}{n_0}\right)^{1/2}
\left(\frac{u}{km s^{-1}}\right)^{-3/2} kpc.
\end{equation}
%--------------------------------------------------------------------

>From this we would like to know the amount of energy required for a 
superbubble to reach the  outer radius of a galaxy with a speed 
comparable to the escape velocity of the system. We shall call this the critical luminosity ($L_{crit}$). Thus, if one sets 
R = R$_G$, u = V$_{esc}$ and the mean number density $n_0$ \si 
$<n_{halo}>$ = $ fM_{ISM}/ \frac{4 \pi}{3} R_{G}^3 \mu_H$; where f is  
the fraction of the galaxy ISM  swept during the 
evolution of the superbubble and  $\mu_H$ is the mean mass per particle,
it leads to:
%--------------------------------------------------------------------
\begin{equation}
      \label{eq.19}
L_{crit,38} = 2.75 \times 10^{-2} \left(\frac{f M_{ISM}}{10^9 M_{\odot}}
\right)\left(\frac{V_{esc}}{km s^{-1}} \right)^3 \left(\frac{R_G}{kpc}
\right)^{-1} .
\end{equation}
%--------------------------------------------------------------------
\noindent
where $L_{crit,38}$ = $L_{crit}$/$10^{38}$ erg.
 
A similar expression can also be found for late evolutionary times 
(t $>$ t$_{burst}$) for which an instantaneous energy release may seem a 
better approximation. Thus, from Chevalier (1974) and Blinnikov et al. (1982)
one obtains:
%--------------------------------------------------------------------
\begin{equation}
      \label{eq.20}
R = 22 \left(\frac{\epsilon E_{50}}{n_0}\right)^{5/21}
t_5^{2/7} \quad pc,
\end{equation}
%--------------------------------------------------------------------
where $E_{50}$ is the total amount of energy deposited by the star cluster
in units of $10^{50}$ erg, and t$_5$ is the evolutionary time in $10^5$ 
years. For the fraction of the total input energy which is transformed into 
the remnant thermal energy $\epsilon = 5/11$ (Weaver et al. 1977). Relation 
(\ref{eq.20}) leads to the critical mechanical luminosity 
$L_{crit} = E_{burst}/t_{burst}$:
%--------------------------------------------------------------------
\begin{equation}
       \label{eq.21}
L_{crit,38} = 8.89 \left(\frac{f M_{ISM}}{10^9 M_{\odot}}
\right) \left(\frac{V_{esc}}{km s^{-1}} \right)^{6/5}.
\end{equation}
%--------------------------------------------------------------------
Both relationships when the fraction of the swept up interstellar mass 
f was set to  0.7  are displayed in Figure 7a,b 
for galaxies with an  
$M_{ISM}$ = 10$^9$ M\sol and $10^{8}$ M\sol, respectively.

 Figure 7a (solid line) shows the $L_{crit}$ trend for 
R$_B$ = 5 kpc and the dashed line  for R$_B$ = 10 kpc (cases A and B). 
The critical luminosity, although by definition it is sufficient to cause the 
ejection of processed matter out of the galaxy, it may only do so,  
depending on the host galaxy mass distribution, after a long evolutionary 
time. In some cases this can be larger than $t_{burst}$ and in fact much
larger than the expected HII region lifetime (\si 10$^7$ yr).
   Several of our numerical calculations in search of $L_{crit}$ are also
shown in  
Figure 7. Diamonds in Figure 7a represent the
numerical results for galaxies with R$_B$ = 5 kpc, and 
triangles  for R$_B$ = 10 kpc models. 
In all cases the time required for the superbubble to reach 
the edge of the galaxy are indicated.

There is a better than 30\% agreement between the derived formulae and the 
numerical results. For superbubbles that require a time 
smaller or  comparable to the starburst time 
t$_{burst}$ to reach the galaxy edge, 
our numerical estimates agree with relation (\ref{eq.19}), and 
for those that take a time longer than $t_{burst}$  
equation (\ref{eq.21}) leads to  an asymptotic  
value of the  
critical energy input rate $L_{crit}$.

An extended halo can therefore inhibit the loss of the swept-up and 
processed matter from a galaxy even if the remnants undergo blowout
from the dense central gaseous component into the halo. 
Note also that the more compact ISM distributions
demand more powerful stellar clusters (a larger $L_{crit}$)
to allow their superbubbles to
reach the outskirts of galaxies and eject their metals into the IGM.
One can therefore conclude that the presence of an extended halo may
dominate the evolution of the remnants, and that the final fate of the
 swept-up galactic gas and of the matter processed by the starburst,
basically depend on the properties of the low density galactic halo 
rather than on the parameters of the gaseous central  disk. Important
is also the extent of the dark matter component: the more compact
this is, the larger the energy required to reach the critical luminosity.

 Note also that all calculations
with an energy input rate 
$\le 10^{41}$ erg s$^{-1}$ and a halo velocity dispersion 
$C_{halo} \geq $ 60 \kms, as inferred from
the observations of amorphous dwarf galaxies (Marlowe etal. 1995, Martin 1996,
etc.), all lie below the curve of critical luminosity, and thus
have led to superbubbles expanding with speeds smaller than the 
random speeds of motions of their host galaxy halos, implying their
disruption long before they reach the edges of their galaxies, and thus
have led to  no loss of their processed matter. Such galaxies in order 
to loose their metals into the intergalactic medium, require 
of energetic starburst, more energetic than what they presently have. 
One could in this respect be tempted to think of a larger energy, 
larger than observed, being deposited in
an extended dwarf galaxy and thus more easely surpase the critical luminosity
value. However, the time required for such superbubbles to reach
the galaxy edge and eject their metals into the IGM would exceed by large
factors the HII region lifetime and thus most probably be    
undetectable events.
 
\section{Conclusions}
 
Based on the scarce but definite observations of amorphous dwarf
galaxies, which show that the HI distribution outgrows by large
factors the optical size of these galaxies, we have constructed 
realistic models which account for an extended cool halo component
bound to the system. Models of amorphous galaxies with an 
M$_{ISM} \sim 10^9 \, M_{\odot}$ and an energy input up to $10^{41}$
 erg  s$^{-1}$, as inferred from the observations, lead to
blowout from the disk matter stratification, allowing the superbubbles 
to vent their hot matter into the galaxy halo.
The outflowing gas immediately searches and follows the steepest
density gradients affecting a large almost spherical volume. We have 
shown here that in the case of small systems, superbubbles  
wrap around and envelope the densest central region of
the host galaxy. Such remnants are strongly decelerated
as they evolve into the halo, sufficiently to remain bound to their 
host galaxy.

Our calculations confirm the results of 
De Young and Gallagher (1990), and De Young and Heckman 
(1994) during the early superbubble evolution.
However, our ultimate conclusions are different as the final
fate of the remnants  basically depends on the properties of the galactic 
halo and on the total energy of the starburst. The presence of the dark
matter component also enhances the ultimate retention of the
processed material. Under these circumstances total retention (i.e. expansion
speeds $\le$ escape velocity) of the remnants and thus of the matter processed
by the exploding massive stars is unavoidable for galaxies with an ISM mass
$\ge$ than a few $10^9$ M$_{\odot}$ and an energy input rate of up to
10$^{41}$ erg s$^{-1}$.

We have also derived analytical expressions for the critical luminosity
required for a superbubble to be able to reach the outskirts of a galaxy
and expulse the processed matter into IGM. The critical limit has
been cross-checked with several numerical calculations 
for a large range of galaxy masses, dimensions and thus 
surface escape velocities.

>From the calculations, we then conclude that the thoroughly studied
dwarf galaxies with a gaseous mass \si 10$^9$ M\sol and presenting   
giant remnants produced by the mechanical
energy ($L \leq 10^{41}$  
erg s$^{-1}$) from massive stars (see Marlowe \etal, 1995; Martin 1996; and
Meurer 1994) are most likely to retain their ISM and their processed matter.
The observed remnants present dimensions smaller than a few  kpc and 
expansion speeds  \si 50 \kms and thus based on our calculations, we predict
 that they have 
already undergone blowout from their disk matter distribution 
and that  presently are 
evolving into the extended low 
density halo of their parent galaxy. Our results also predict that these
superbubbles will be ultimately dispersed by the turbulent motions in the 
extended galactic halo.

The consequences of these conclusions are important in a variety of fields
and we shall analyse these in a forthcomming communication.
Clearly, the kind of dwarf galaxies that we have analysed, are not responsible 
for the contamination of the IGM. Dwarf galaxies with an ISM mass 
of the order of 10$^9$ M\sol will retain their metals unless they undergo
an overwhelming burst of star formation much larger than those presently 
observed. We thus have to understand why do they present low metalicities. 
The secret must probably lies on the physics of mixing (see \GTT 1996).

\acknowledgements

%{\bf Acknowledgments}

We thank our anonymous referee for many suggestions that allow us to
revise and improve our knowledge of dwarf galaxies.
SAS acknowledges support from the Royal Society grant for joint projects
with the former Soviet Union, and the RGO and the IoA in 
Cambridge for partial financial support. GTT gratefully acknowledges the BBV 
foundation for a visiting professorship at the University of Cambridge. We 
also thank the hospitality and friendly assistance of the
Instituto Nacional de Astrof\'\i{}sica Optica y Electr\'onica during our
visit to Mexico.  It is also our pleasure to thank G.S. Bisnovatyi-Kogan 
for many discussions on the blowout phenomena, and Casiana Mu\~noz Tu\~non 
and Evan Skillman for useful suggestions. We also thank E. Terlevich
and R. Terlevich for their suggestions and a careful reading of the
manuscript. We also thank Dr. A. Suchkov who provided us with a table with 
specific X-ray emissivities. 

\clearpage

--------------------------------------------------------------------
\setcounter{equation}{0}
\renewcommand{\theequation}{A\arabic{equation}}
{\center \section*{\bf Appendix A: Main equations }}

Following TI, Li \& Ikeuchi (1992), 
TB, SBHL our approach includes three co-existing ISM 
phases: a neutral, an ionized component, and a low density extended 
halo characterized by a  turbulent velocity; also thought
by other authors as a collection of low densities clouds with a 
velocity dispersion $C_{halo}$.

{\bf Mass conservation.} The expanding shell caused by the star cluster energy deposition accumulates interstellar mass according to
%-----------------------------------------------------
\begin{equation}
      \label{a.1}
{{\rm d}\mu_L\over {{\rm d}t}} = {{\rm d}\mu_g \over {{\rm d}t}} -
%{{\rm d}\mu_{ic}\over {{\rm d}t}} + \der{<\mu_{CM}>}{t} -
       {{\rm d}\mu_{RT} \over {{\rm d}t}} -
       {{\rm d}\mu_{cusp} \over {{\rm d}t}},
\end{equation}
%------------------------------------------------------------
where the first term represents the mass growth due to swept-up
interstellar gas, the second one the loss of mass resultant from the
development of Rayleigh-Taylor (RT) instabilities after the shell blows out 
of the galaxy disk, and the third one describes possible shell self 
intersections following cusp formation (see below). 
We have assumed  that the swept turbulent halo sticks to the shell. 
Then for any Lagrangian element with a surface area ${\rm d}\Sigma$ and a 
unit vector normal to the shell ${\bf n}_i$, the terms in equation 
(\ref{a.1}) take the form:
%--------------------------------------------------------------
\beqa
      \label{a.2}
      & & \hspace{-0.5cm}
\der{\mu_g}{t} = \der{\mu_{ic}}{t} + \der{<\mu_{halo}>}{t}
  \\[0.2cm]
      & & \hspace{-0.5cm}
{{\rm d}\mu_{ic}\over {{\rm d}t}}  = \left\{
\begin{array}{lcl}
(\rho_{NM}+\rho_{IM})({\bf u}_i-{\bf v}) \cdot {\bf n}_i {\rm d}\Sigma,
\ \ \  u_n \ge a_s \\ [0.2cm]
0, \ \ \ 
u_n < a_s \\ 
\end{array}
\right.
\eeqa
%----------------------------------------------------------------
\beq
      \label{a.3}
\der{<\mu_{halo}>}{t} = <\rho_{halo}>({\bf u}_i-{\bf v}_{halo}) \, 
{\bf n}_i {\rm d}\Sigma, 
\eeq
%---------------------------------------------------------------------
\begin{equation}
      \label{a.4}
{{\rm d}\mu_{RT}\over {{\rm d}t}}  = \left\{
\begin{array}{lcl}
 0, \ \ \  a_{sh} < 0 \\ [0.2cm]
 \frac{\mu_L(t=t_{RT})}{\tau_{RT}} \ \ \  \\ [0.2cm]
 0, \ \ \  \mu_{RT} > \mu_L(t=t_{RT}) \\
\end{array}
\right.
\end{equation}
%-----------------------------------------------------
\begin{equation}
      \label{a.5}
{{\rm d}\mu_{cusp}\over {{\rm d}t}}  = \delta(t-t_{cusp}) \mu_{L,rem}
\end{equation}
%---------------------------------------------------------------------
An effective sound velocity $a_s=(A_n C_{NM}^2 + A_i C_{IM}^2)^{1/2}$,
where $A_n =\rho_{NM}/(\rho_{NM} + \rho_{IM})$, $A_i=\rho_{IM} /
(\rho_{NM} + \rho_{IM})$, $a_{sh}$ is the normal component of the shell
acceleration. $t_{cusp}$ is the time of cusp formation, $t_{RT}$ marks
the time for the beginning of the acceleration phase, $\mu_L(t=t_{RT})$ 
is the mass of the Lagrangian mesh at the beginning of blow-out, and
$\tau_{RT}$ is the characteristic growth time for RT instabilities
%-----------------------------------------------------
\begin{equation}
      \label{a.6}
\tau_{RT} = \left(\frac{\lambda}{2 \pi a_{sh}} \right)^{1/2}
\end{equation}
%------------------------------------------------------------
where $\lambda$ is the same order of magnitude as the characteristic
scale for the ambient gas density distribution, $\lambda \approx 2H_z$
(Koo \& McKee, 1990). For simplicity we set here $\lambda$ equal to the
characteristic scale for the gas density distribution at the
galactic center $\lambda=H_{z0}$.
% (for the calculation of $H_{z0}$ see Appendix C).

{\bf Momentum conservation.}
We consider four effects leading to changes in the shell momentum: the
pressure difference between the inside gas  and the surrounding medium 
$\Delta P$, the galactic gravitational force ${\bf g}$, the accumulation 
of the ambient gas momentum, and the momentum loss during the RT break-out. 
We assume that RT clumps separate from the shell with the same speed as 
the current shell expansion velocities. Also that intercloud mass 
accumulation stops wherever the expansion velocity decreases below the 
sound speed $a_s$, and thus the interaction with the ambient intercloud gas 
modifies only the normal component of the momentum but the shell does not 
accumulate mass any further. With these simplifications, the equation 
of motion for any Lagrangian element with mass $\mu_L$ is (see also 
Ostriker \& McKee 1988, Silich et al. 1996):
%--------------------------------------------------
\begin{eqnarray}
      \label{a.7}
      & & \hspace{-0.5cm}
 {{\rm d}(\mu_L {\bf u})\over {{\rm d}t}} = \Delta P{\bf n}{\rm d}\Sigma
      + {\bf v}{{\rm d}\mu_{ic} \over {\rm d}t} 
      +{\bf v}_{halo} \der{\mu_{halo}}{t} -
 {\bf u}{{\rm d}\mu_{RT}\over {{\rm d}t}}  +  
   \nonumber \\[0.2cm]
      & & \hspace{-0.5cm}
+ \mu {\bf g} - \rho_{ic} (x,y,z)[({\bf u}- {\bf v}) \cdot {\bf n}_i]^2
 {\bf n} {\rm d}\Sigma
\end{eqnarray}

%--------------------------------------------------
Then the equation for the  shell expansion velocity can be written as 
%--------------------------------------------------
\begin{eqnarray}
      \label{a.8}
      & & \hspace{-0.5cm}
 {{\rm d}{\bf u}\over {{\rm d}t}} = \mu_L^{-1}\Delta P{\bf n}{\rm d}\Sigma
      + {\bf (v-u)} \mu_L^{-1}{{\rm d}\mu_{ic} \over {\rm d}t}
      + ({\bf v}_{CM}- {\bf u}) \mu_L^{-1}{{\rm d}\mu_{halo} \over {\rm d}t} +
   \nonumber \\[0.2cm]
      & & \hspace{-0.5cm}
+ {\bf g} - \rho_{ic} (x,y,z) \mu_L^{-1}[({\bf u}- {\bf v}) \cdot {\bf
n}]^2 {\bf n} {\rm d}\Sigma,
      \\[0.2cm]
      & & \hspace{-0.5cm}
      \label{a.9}
 {{\rm d}{\bf r}_i\over {{\rm d}t}} = {\bf u}
\end{eqnarray}
%--------------------------------------------------
In the current calculations we have assumed that the intercloud gas
and halo clouds move with the same regular velocities 
($\bf{v}$ = ${\bf v}_{halo}$).

{\bf The energy balance.}
The total energy balance for a remnant caused by the strong energy 
input from the central star cluster reads as:
%-----------------------------------------------------
\begin{equation}
      \label{a.23}
E_{SN} + E_{k,g} + E_{th,g} = E_{th} + E_{k,sh} + E_{th,sh} + E_g +
E_{r,in} + E_{r,sh} + E_{k,RT},
\end{equation}
%------------------------------------------------------------
where the left-hand terms are the energy deposited by massive stars 
($E_{SN}$), and the initial 
kinetic and thermal energies of the accumulated ISM 
($E_{k,g}$ and $E_{th,g}$). The right-hand terms are the
bubble interior thermal energy ($E_{th}$), the kinetic and the thermal
energies of the shell ($E_{k,sh}$ and $E_{th,sh}$), the shell gravitational 
energy ($E_g$), the energy losses from the hot bubble interior ($E_{r,in}$) 
and from the shell ($E_{r,sh}$), and the  kinetic energy of RT clumps
($E_{k,RT}$).
The derivative of the bubble thermal energy could be expressed as follows
%----------------------------------------------------------------------
\begin {equation}
       \label{a.10}
\frac{{\rm d} E_{th}}{{\rm d}t} = L_{burst} - L_{in} -
P_{in} \frac{{\rm d} \Omega}{{{\rm d}t}},
\end {equation}
%----------------------------------------------------------------------
where
%----------------------------------------------------------------------
\begin{equation}
       \label{a.11}
\frac{{\rm d} \Omega}{{{\rm d}t}} = \sum_{i=1}^{N} u_n
{\rm d} \Sigma ,
\end{equation}
%---------------------------------------------------------------------
$L_{in}$ is the energy loss from the hot bubble interior, and $u_n$ is
the normal component of the expansion velocity. As the main contribution
for the interior bubble cooling comes from the outer dense layers, we
approximate $L_{in}$ as  
%-------------------------------------------------------------------
\begin {equation}
       \label{a.28}
L_{in} = L_R \frac{\Sigma^R}{\Sigma}=\varepsilon_2 L_R,
\end {equation}
%-----------------------------------------------------------------
where $\Sigma$ is the total surface of the remnant,and the index R denotes 
the radiative
part of the shell, $\varepsilon_2 = \Sigma^R / \Sigma$. $ L_R$
is the energy loss from the bubble interior if  all the remnant becomes
radiative (for the calculations of $ L_R$ see Silich et al. 1996).

One can then calculate the bolometric shell luminosity from the total energy
balance: 
%----------------------------------------------------------------------
\begin{equation}
       \label{a.29}
L_{sh} = P_{in}\der{\Omega}{t} + 
\der{E_{k,g}}{t} + \der{E_{th,g}}{t} 
- \der{E_{th,sh}}{t} - 
\left(\frac{{\rm d}E_{k,sh}}{{\rm d}t} +
\frac{{\rm d}E_{k,RT}}{{\rm d}t} \right) - 
\frac{{\rm d}E_g}{{\rm d}t},
\end{equation}
%-----------------------------------------------------------
where 
%--------------------------------------------------------------
\begin{eqnarray}
       \label{a.25}
      & & \hspace{-0.5cm}
\frac{{\rm d} E_{k,sh}}{{\rm d}t} = \frac{1}{2}
\sum_{i=1}^{N} \left({\dot \mu}_L u^2 + 2 \mu_L {\bf u}
{{\rm d}{\bf u} \over {{\rm d}t}} \right)
      \\[0.2cm]
      & & \hspace{-0.5cm}
            \label{a.26}
{{\rm d} E_{k,g} \over {{\rm d}t}}  = \left\{
\begin{array}{lcl}
 \frac{1}{2} \sum_{i=1}^{N} ({\dot \mu_{ic}}+ {\dot \mu_{halo}}) v^2, \ \ \
 u_n \ge a_s \\ [0.2cm]
\frac{1}{2} \sum_{i=1}^{N} ({\dot \mu_{ic}} (V_n^2 - u_n^2) +
{\dot \mu_{halo}} v^2), \ \ \ u_n < a_s   \\
\end{array}
\right.
      \\[0.2cm]
      & & \hspace{-0.5cm}
            \label{a.26a}
\der{E_{th,g}}{t} = \sum_{i=1}^{N} \frac{{\dot \mu_g}}
{\gamma(\gamma-1)} 
\left[\frac{\rho_{NM}}{\rho_g} C_{NM}^2 + 
\frac{\rho_{IM}}{\rho_g}C_{IM}^2 
+ \frac{\gamma(\gamma-1)}{2} \frac{
\rho_{halo}}{\rho_g} C_{halo}^2 \right]
      \\[0.2cm]
      & & \hspace{-0.5cm}
            \label{a.27a}
\frac{{\rm d} E_g}{{\rm d}t} = - \sum_{i=1}^{N} \mu_L {\bf u} {\bf g}
      \\[0.2cm]
      & & \hspace{-0.5cm}
            \label{a.27b}
\frac{{\rm d} E_{k,RT}}{{\rm d}t} =  \frac{1}{2} \sum_{i=1}^{N_{RT}}
{\dot \mu}_{RT} u^2,
\end{eqnarray}
%---------------------------------------------------------------------

(\ref{a.29}) gives a good estimate of the radiative shell luminosity if 
one sets
 
%------------------------------------------------
\beq
       \label{a.x}
\der{E_{th,sh}}{t} = 0 .
\eeq
%-----------------------------------------------------
However, in the thin layer approximation there is no satisfactory 
procedure for the calculation of an adiabatic shell thermal energy. 
Therefore for quasi-adiabatic shell sectors we use:
%--------------------------------------------------------------------
\begin{equation}
      \label{b.2}
L_{sh}^A =  \xi \sum_{i=1}^{N_A} n_{sh,a}^2 \Lambda_t(T_{sh,a}) \Delta R
{\rm d} \Sigma^A,
\end{equation}
%---------------------------------------------------------------------
where $\xi$ is the galactic gas metallicity, $\Lambda_t(T)$ is cooling
function, 
$n_{sh,a}$ and $T_{sh,a}$ are the mean atomic particle number density
and temperature within the adiabatic part of the shell, $\Delta R = 0.14 R_s$
is the quasi-adiabatic shell thickness, ${\rm d} \Sigma^A$ is the surface
area of the adiabatic Lagrangian mesh, and $N_A$ is the number of the
"adiabatic" Lagrangian meshpoints.
%--------------------------------------------------------------------
\begin{eqnarray}
      & & \hspace{-0.5cm}
      \label{b.3}
n_{sh,a} = \frac{\mu_L}{\mu_{in} \Delta R {\rm d} \Sigma^a}
      \\[0.2cm]
      & & \hspace{-0.5cm}
            \label{b.4}
T_{sh,a} = \frac{1}{k} \frac{P_{in}}{n_{sh,a}}
\end{eqnarray}
%---------------------------------------------------------------------
The total shell luminosity then follows from the expression
%--------------------------------------------------------------------
\begin{equation}
      \label{b.5}
L_{sh} = L_{sh}^a +  L_{sh}^R,
\end{equation}
%---------------------------------------------------------------------
whereas the bolometric bubble luminosity is 
%------------------------------------------------------------------
\begin{equation}
      \label{a.45}
L_{bol} = L_{in} + L_{sh}, 
\end{equation}
%---------------------------------------------------------------------

\clearpage
\setcounter{equation}{0}
\renewcommand{\theequation}{B\arabic{equation}}
{\center \section*{\bf Appendix B: The inner bubble structure}}

Our approximate description of the bubble structure is based 
on the similarity solution of Weaver et al. (1977). We assume that the 
density and temperature distribution can  be approximated by:
%--------------------------------------------------------------------
\begin{eqnarray}
      \label{a.32}
      & & \hspace{-0.5cm}
n = n_c (1 - x_i)^{-\lambda},
      \\[0.2cm]
      \label{a.33}
      & & \hspace{-0.5cm}
T = T_c (1 - x_i)^{\lambda},
\end{eqnarray}
%--------------------------------------------------------------------
where $x_i = r / R_i$ is the dimensionless distance from the cavity center 
to the particular Lagrangian element i, and $\lambda = 2/5$. To determine
$n_c$ (and $T_c$) we assume that the total mass interior to superbubble
at all times is:
%--------------------------------------------------------------------
\begin{equation}
      \label{a.33a}
M_{tot} = M_{SN} + M_{ev},
\end{equation}
%--------------------------------------------------------------------
where 
%--------------------------------------------------------------------
\begin{eqnarray}
      \label{a.33b}
      & & \hspace{-0.5cm}
M_{SN} = \int_o^{t_{burst}} {\dot M}_{SN} \dif t , 
      \\[0.2cm]
      \label{a.33c}
      & & \hspace{-0.5cm}
M_{ev} = \int_o^{t}{\dot M}_{ev} \dif t .  
\end{eqnarray}
%--------------------------------------------------------------------
The value of ${\dot M}_{SN}$ is defined by (\ref{eq.15}), and the 
shell thermal evaporation  is (Mac Low \& McCray, 1988): 
%------------------------------------------------------------------------
\begin{equation}
%      & & \hspace{-0.5cm}
      \label{a.43}
\dot M_{ev}  = \left\{
\begin{array}{lcl}
\frac {4}{25} \frac{\mu_{in}}{k}C{T_c}^
{5/2} \sum_{i=1}^{N_R} \frac {d {\Sigma}_i}{R_i},
\ \ \  radiative \quad  phase,   \\ [0.2cm]
0, \ \ \ 
adiabatic \quad phase,       \\ 
\end{array}
\right.
%\dot M_{ev}=\frac {4}{25} \frac{\mu_{in}}{k}C{T_c}^
%{5/2} \sum_{i=1}^{N_R} \frac {d {\Sigma}_i}{R_i}, at the radiative phase,
%   \\[0.2cm]
%      & & \hspace{-0.5cm}
%\dot M_{ev}= 0, at the adiabatic phase,
\end{equation}
%--------------------------------------------------------
where  $k$ is the Boltzmann's constant, $C=6 \times 10^{-7}$ erg 
s$^{-1}$ cm$^{-1}$ K$^{-2/7}$ is the constant  thermal
conductivity coefficient for a fully ionized hydrogen plasma 
(Cowie \& McKee 1977).

Two approximations are then used. We assume that a portion of the total mass
($M_1$) is enclosed within the radius of the contact discontinuity 
($R_{CD} = \varepsilon_1 R_s$)
%--------------------------------------------------------------------
\begin{equation}
      \label{a.34}
M_1 = \mu_{in} n_c \int_0^{2 \pi} {\rm d} \phi \int_0^{\pi} \sin \theta
         {\rm d} \theta \int_0^{R_{CD}} (1-r/R_s)^{-\lambda}r^2 
{\rm d}r = 3 \mu_{in} n_c I_{m1}(\lambda) \Omega,
\end{equation}
%--------------------------------------------------------------------
\begin{eqnarray}
      & & \hspace{-0.5cm}
      \label{a.35}
I_{m1}(\lambda) = \int_0^{\varepsilon_1} (1 - x)^{-\lambda} x^2 {\rm d} x =
\frac{1}{1-\lambda} \left[1 -(1-\varepsilon_1)^{1-\lambda} \right]
   \nonumber \\[0.2cm]
      & & \hspace{-0.5cm}
+ \frac{1}{3-\lambda} \left[1 -(1-\varepsilon_1)^{3-\lambda} \right]
- \frac{2}{2-\lambda} \left[1 -(1-\varepsilon_1)^{2-\lambda} \right].
\end{eqnarray}
%--------------------------------------------------------------------
$M_1 = M_{tot}$ during the adiabatic phase but as sectors of the shock
become radiative, $M_{tot}$ fills a larger volume. In such case, the fraction 
of $M_{tot}$ that  now reaches up to $R_s$, assuming a collapsed thin 
shell with radius equal to that of the shock, is simply 
%--------------------------------------------------------------------
\begin{equation}
      \label{a.36}
M_2 =   \varepsilon_2 (M - M_1),
\end{equation}
%--------------------------------------------------------------------

%--------------------------------------------------------------------
\begin{equation}
      \label{a.37}
M = 3 \mu_{in} n_c I_m(\lambda) \Omega,
\end{equation}
%--------------------------------------------------------------------
\begin{equation}
      \label{a.38}
I_m(\lambda) = \int_0^1 (1 - x)^{-\lambda} x^2 {\rm d} x =
\frac{1}{1-\lambda} + \frac{1}{3-\lambda} - \frac{2}{2-\lambda}.  
\end{equation}
%--------------------------------------------------------------------

The mass distribution of the (adiabatic-radiative) remnant interior 
can be defined as 
%--------------------------------------------------------------------
\begin{equation}
      \label{a.39}
M_1 + M_2 = 
\varepsilon_2 M + (1 - \varepsilon_2) M_1 = M_{tot} . 
\end{equation}
%--------------------------------------------------------------------
Equations (B3 - B12)  define the central superbubble density 
and temperature:
%--------------------------------------------------------------------
\begin{eqnarray}
      \label{a.40}
      & & \hspace{-0.5cm}
n_c = \frac{1}{3 \mu_{in} [\varepsilon_2 I_m(\lambda) + (1-\varepsilon_2)
I_{m1}(\lambda)]} \frac{M_{tot}}{\Omega},
      \\[0.2cm]
      \label{a.41}
      & & \hspace{-0.5cm} 
T_c = \frac{3 \mu_{in}[\varepsilon_2 I_m(\lambda) + (1-\varepsilon_2)
I_{m1}(\lambda)]}{k} \frac{P_{in} \Omega}{M_{tot}}.
\end{eqnarray}

\subsubsection{X-ray luminosity}

The total bubble X-ray luminosity includes the X-ray emission from the hot
bubble interior and the X-ray emission of the swept-up interstellar
gas. We conventionally divide the X-ray emission from the bubble interior
into an emission $L_{x1}$, from the matter within the radius $R_{CD}$, and 
that arising from the gas between the 
"adiabatic CD" surface and the X-ray cutt-off surface, $L_{x2}$. The 
first part is defined by the equation
%--------------------------------------------------------------------
\begin{equation}
      \label{a.46}
L_{x1} = \int_0^{2 \pi} {\rm d} \phi \int_0^{\pi} \sin \theta {\rm d} \theta
  \int_0^{\varepsilon_1 R_s} \xi n^2_a(r) \Lambda_x r^2 {\rm d}r = 
  3 \xi \lambda^{-1} n^2_{ac} \Omega I_{x1}
\end{equation}
%---------------------------------------------------------------------
where
%--------------------------------------------------------------------
\begin{equation}
      \label{a.47}
I_{x1} = \frac{1}{T_c} \int_{T_{CD}}^{T_c} \Lambda_x (T) 
\left(\frac{T}{T_c} \right)^{\frac{1-3\lambda}{\lambda}}
\left[1 - \left(\frac{T}{T_c} \right)^{\frac{1}{\lambda}}\right]^2 
{\rm d} T
\end{equation}
%---------------------------------------------------------------------
The second part we approximate by the value 
%--------------------------------------------------------------------
\begin{equation}
      \label{a.48}
L_{x2} = (L_{xr} - L_{x1}) \varepsilon_2, 
\end{equation}
%---------------------------------------------------------------------
where $L_{xr}$ is the X-ray emission from the interior of the fully 
radiative bubble (see Silich et al. 1996):
%--------------------------------------------------------------------
\begin{eqnarray}
      \label{a.49}
      & & \hspace{-0.5cm}
L_{xr} = 
\int_0^{2 \pi} {\rm d} \phi \int_0^{\pi} \sin \theta {\rm d} \theta
  \int_0^{R_s} \xi n^2_a(r) \Lambda_x r^2 {\rm d}r = 
  3 \xi \lambda^{-1} n^2_{ac} \Omega I_{xr},
      \\[0.2cm]
     & & \hspace{-0.5cm} 
I_{xr} = 
\frac{1}{T_c} \int_{T_{x,cut}}^{T_c} \Lambda_x (T) 
\left(\frac{T}{T_c} \right)^{\frac{1-3\lambda}{\lambda}}
\left[1 - \left(\frac{T}{T_c} \right)^{\frac{1}{\lambda}}\right]^2 
{\rm d} T.
\end{eqnarray}
%---------------------------------------------------------------------
Here $T_{CD}$ is the temperature at the "adiabatic CD" surface, and 
$T_{x,cut} \approx 5 \times 10^5K$ is the X-ray cut-off temperature 
(see Chu \& Mac Low, 1990; also SBHL). 
The X-ray luminosity of the hot bubble interior is then
%--------------------------------------------------------------------
\begin{equation}
      \label{a.50}
L_{x,in} =  \varepsilon_2 L_{xr} + (1 -\varepsilon_2) L_{x1}. 
\end{equation}
%---------------------------------------------------------------------
The X-ray luminosity from the shell includes both the X-ray emission from
the quasi-adiabatic sectors of the shell $L_{x,sh}^a$ and the X-ray
emission from the outer layers of the radiative shell $L_{x,sh}^R$. 
We approximate $L_{x,sh}^a$ by the expression for the homogeneous
shell luminosity 
%--------------------------------------------------------------------
\begin{equation}
      \label{a.51}
L_{x,sh}^A =  \xi \sum n_{sh,a}^2 \Lambda_x(T_{sh,a}) \Delta R 
{\rm d} \Sigma^A, 
\end{equation}
%---------------------------------------------------------------------
and use the expression
%----------------------------------------------------------------------
\begin{equation}
      \label{a.53}
L_{x,sh}^R  = \left\{
\begin{array}{lcl}
 0, \ \ \   T_s < T_{x,cut}, \\ [0.2cm]
\sum_{i=1}^{N_R} \left(\frac{T_{max}}{T_s} \right)^2
\frac{\Lambda_x(T_s)} {\Lambda_{tot}(T_{max})} L_{b,sh}^R,  
\ \ \   T_s \ge T_{x,cut} ,  \\
\end{array}
\right.
\end{equation}
%-------------------------------------------------------------------
to estimate the X-ray luminosity from the outer shock front. Here 
the mean shell density and temperature are defined by formulae
(\ref{b.3}) - (\ref{b.4}), ${\rm d} \Sigma^a$ is the surface area of the 
adiabatic Lagrangian mesh, $T_{max} \approx 10^5K$ is the temperature 
at the peak of the cooling function $\Lambda_{tot}$.
The shell X-ray emission then equals
%--------------------------------------------------------------------
\begin{equation}
      \label{a.54}
L_{x,out} = L_{x,sh}^A + L_{x,sh}^R
\end{equation}
%---------------------------------------------------------------------
and the total bubble X-ray luminosity reads as 
%--------------------------------------------------------------------
\begin{equation}
      \label{a.55}
L_x = L_{x,in} + L_{x,out}.
\end{equation}
%---------------------------------------------------------------------

%\subsubsection{Post-shock parameters}
%
%Since our numerical scheme does not include full description for the
%cloud-shell interaction, we assume simply that clouds stick with a
%shell and convert energy of the random motion into the shell thermal
%energy. One will be able then to estimate the post-shock temperature
%and density from the ordinary relations
%--------------------------------------------------------------------
%\begin{eqnarray}
%      & & \hspace{-0.5cm}
%      \label{a.56}
%n_s = \frac{(\gamma+1) M_s^2}{(\gamma+1) M_s^2 + 2} \rho_g
%      \\[0.2cm]
%      & & \hspace{-0.5cm} 
%T_s = \frac{[2 \gamma M_s^2 - (\gamma-1)][(\gamma-1)M_s^2 +2]}
%{(\gamma + 1)^2 M_s^2} T_{eff},
%\end{eqnarray}
%---------------------------------------------------------------------
%if introduce an effective ambient medium temperature 
%--------------------------------------------------------------------
%\beq
%T_{eff} = \frac{1}{k} \frac{P_{ext}}{n_g},
%\eeq
%--------------------------------------------------------------------
%and sound velocity which for the cloudy component is equivalent to the
%assumed velocity dispersion (SBHL).

\subsubsection{Phase transitions}

During the evolution in a disk-halo system, superbubbles undergo several 
transitions (from quasi-adiabatic to radiative and viceversa). The time for
phase transition can be estimated by comparing the characteristic 
cooling time
%--------------------------------------------------------------------
\begin{equation}
      \label{a.57}
\tau_{cool} = \frac{3 \mu_a}{2 \mu_{in}} \frac{kT}{\xi n_a \Lambda_{tot}}.
\end{equation}
%---------------------------------------------------------------------
with the dynamic time (t).  Here $n_a$ is the atomic number density and
the ratio of $\mu_a / \mu_{in}$ takes into account gas ionization. We
assume the mean shell density and temperature (see Appendix A) 
to estimate the time for the adiabatic - radiative transition, and use
Gaetz and Salpeter (1983) cooling function up to $4 \times 10^7K$.
The $T^{1/2}$ growth of the cooling rate for higher temperatures is assumed.
The sudden loss of the thermal energy during the
adiabatic to radiative transition (see Chevalier 1974; and Falle 1981),
is taken into account and is assumed to be in a direct proportion to the 
surface of the rapidly cooling part of the shell:
%--------------------------------------------------------------------
\begin{equation}
      \label{a.58}
\Delta E_{th} = \frac{\varepsilon_2}{2}E_{th}.
\end{equation}
%---------------------------------------------------------------------
Thus, if the adiabatic-radiative transition occurs for a spherical
shell, the remnant looses half it's current thermal energy.

\subsubsection{Cusps formation and grid rearrangement}

As the shell reaches several characteristic scales $H_z$, the acceleration
phase begins, and the shell experiences Rayleigh-Taylor instabilities.
Once the accelerated sector of the shell is disrupted, a secondary
shell forms in the low density halo. Soon the bottom parts of
the "cap" begin to overlap and intersect the vertically elongated main
body of the remnant and a ring-like cusp forms. Some of the Lagrangian
particles then move into the bubble interior what makes the later
calculations impossible. To avoid this difficulty
we have introduced a simple geometric criterion. The cylindrical
radius of any Lagrangian element at any Lagrangian $"\theta"$-slice (with 
the exception of top, bottom, and those, which have  maximum radii) should
be either greater than the cylindrical radius of the previous Lagrangian 
mesh element and smaller than the radius of the next one, or it should be 
smaller and greater than the radii of the previous and following Lagrangian 
mesh particles. If this is not the case, we assume cusp
formation and the mass of the corresponding particle goes into the
remnant interior and is not considered any further.  

We also exclude from future consideration those
Lagrangian zones which evolve and interpenetrate the galaxy midplane. 
To maintain a sufficient space resolution we regrid the shell,
if the total number of Lagrangian zones, which are parallel to the 
galaxy plane falls below the initial value $N_z$.

\clearpage

{\bf Figure captions}

1) The ISM of dwarf galaxies. a-d show the density distribution
for models A120, A100, A80, and A60. For comparison purpose e
shows model B60 and f shows the distribution obtained for model
C50 (see Table 1). In all frames the outer contour was drawn at
the edge of the galaxy. 

2) The ISM mass distribution. The distribution of the 10$^{9}$
M\sol and 10$^{8}$ M\sol ISM for models A and C (see Table 1),
respectively. 

3) The time evolution. Evolutionary sequence showing the shape and
development of the superbubble for the cases A100 and B60 (see
Table 1). The various contours are drawn at t = 14 $\times 10^6$
yr, 26 $\times 10^6$ yr and 50 $\times 10^6$ yr. 

4) The time evolution. Panel a displays the amount of ISM swept
by the superbubble calculated for the case A100 (see Table 1), as
a function of time, while b and c display the matter deposited
into the superbubble interior by thermal evaporation from the
shell of swept up matter and that deposited by the sequential SNe.
d and e present the calculated bolometric and X-ray luminosity 
of the superbubble. The latter one shows the contribution from 
the swept up shell being much smaller than that produced by the 
remnant interior which accounts for most of the total X-ray
luminosity.  

5) The fate of processed matter. Panels a and b compare the maximum
expansion speed, measured along the symmetry axis (solid lines),
with the escape velocity as a function of distance to the centre
of the galaxy (dashed lines), for all A and B cases. The dotted
lines mark the moment when the fastest expansion speed drops below
the assumed halo velocity dispersion.   

6) The fate of processed matter. The ratio of the pressure within
the superbubble (for models A) to that at the edge of its host
galaxy, as a function of the superbubble size. The sudden change
in slope is , in all cases, due to the sudden cut in energy
injection at $t_{burst}$ = 40 $\times 10^{6}$ yr. The dotted line 
indicates the run of the superbubble interior pressure for the model 
A60 normalized to the 
ISM pressure at the shell stall radius.

7) The critical luminosity. Panels a and b show the critical luminosity
values (derived from eq 26; solid lines) required for a
superbubble evolving into a $M_{ISM}$ = 10$^9$ M\sol (a) or 
 $M_{ISM}$ = 10$^8$ M\sol (b) to reach the edge of its host galaxy
with a velocity similar to the escape velocity measured 
at the edge of the system. Larger luminosities would produce 
faster remnants and thus their processed matter would be lost from
the galaxy. Panel (a) also shows the luminosity values obtained
from the same eq 26 (dashed line) for all B cases; \ie the cases
for which the dark matter component is assumed to be distributed
within a radius of 5 kpc, instead of 10 kpc. The dotted line (in a
and b) indicates the luminosity values derived from eq 28,
regarded as a better approximation when the evolutionary time
exceeds the $t_{burst} = 40 \times 10^6$ yr. The diamond and
triangular symbols indicate the critical luminosity values derived
from our numerical simulations, after a number of iterations. 
For each of these, the indicated ages mark the evolutionary
time required for the superbubble to reach the edge of the system.       
Those exceeding  $t_{burst}$ should then be compared with the 
solution represented with the dotted line.

\clearpage

{\singlespace
\small

\centerline {\bf {Table 1}}
\centerline {Model parameters}
\scriptsize
\vspace{0.4cm}
%\hskip -0.7truein
\begin{tabular} {lccccccccc} \hline
& $ M_{tot} $ & $ M_{gas} $ &  $ R_B $ & 
$ R_{G} $ & $ C_{halo} $ & $ n_{0} $ & $ E_{burst} $ & $ L $ 
& Shell evap.  \\
N & $10^9 M_{\odot}$ & $10^9 M_{\odot}$ & kpc 
& kpc &km/s$^{-1}$ &cm$^{-3}$ &$10^{55}$ ergs &$10^{40}$erg s$^{-1}$& \\ 
\hline
A120 & 10.0 & 1.0 & 10 & 3.9  & 120 & 20.3 &10 & 7.9 & yes \\
A100 & 10.0 & 1.0 & 10 & 8.5  & 100 & 20.2 &10 & 7.9 & yes \\
A80  & 10.0 & 1.0 & 10 & 13.6 &  80 & 20.2 &10 & 7.9 & yes \\ 
A60  & 10.0 & 1.0 & 10 & 24.0 &  60 & 20.2 & 5 & 3.95 & yes \\ 
$A^{\star}$100 & 10.0 & 1.0 & 10 & 8.5  & 100 & 20.2 &10 & 7.9 & no \\
\hline
B160 & 10.0 & 1.0 &  5 & 2.8  & 160 & 20.6 &10 & 7.9 & yes \\ 
B120 & 10.0 & 1.0 &  5 & 6.0  & 120 & 20.3 &10 & 7.9 & yes \\ 
B100 & 10.0 & 1.0 &  5 & 8.7  & 100 & 20.3 &10 & 7.9 & yes \\ 
B60  & 10.0 & 1.0 &  5 & 24.0 &  60 & 21.6 & 5 & 3.95 & yes \\ 
\hline
C50 & 1.0 & 0.1 & 5 & 3.0  & 50 & 20.05 & 1.07 & 0.85 & yes \\ 
C40 & 1.0 & 0.1 & 5 & 5.4  & 40 & 20.20 & 0.57 & 0.45  & yes \\
C30 & 1.0 & 0.1 & 5 & 9.7  & 30 & 20.20 & 0.42 & 0.33  & yes \\
\hline
D80 & 5.0 & 1.0 & 10 & 5.7  & 80 & 20.26 & 10 & 7.9 & yes \\ 
E80 & 5.0 & 1.0 &  5 & 6.8  & 80 & 20.36 & 10 & 7.9 & yes \\
\end{tabular}
%----------------------------------------------------------------------

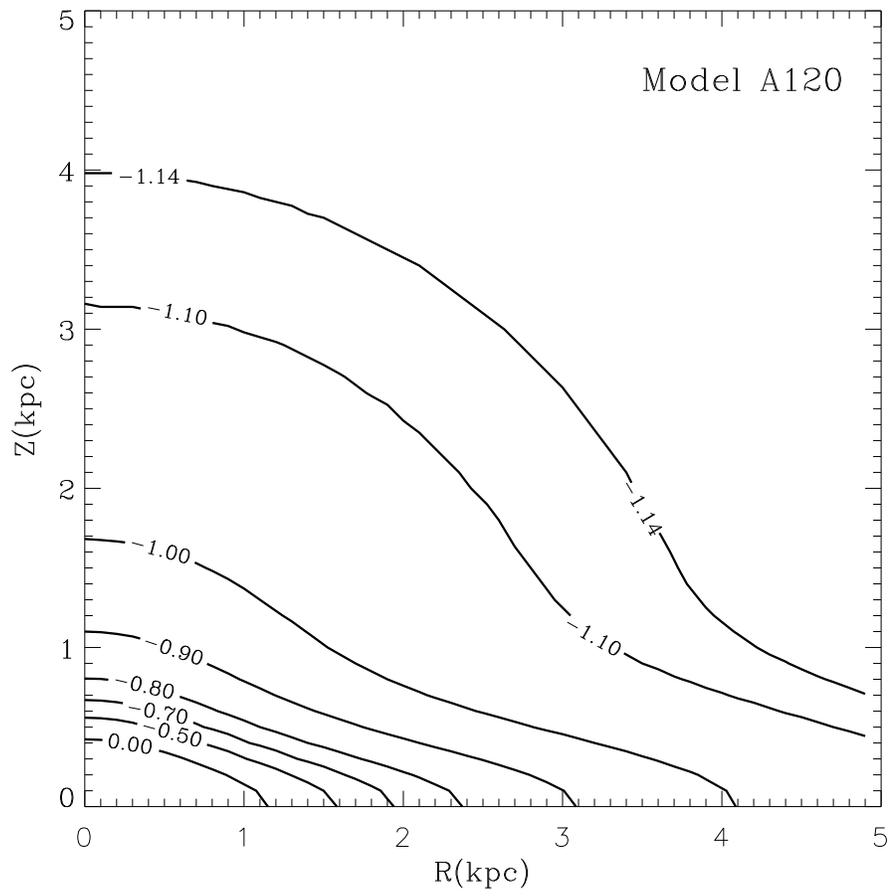
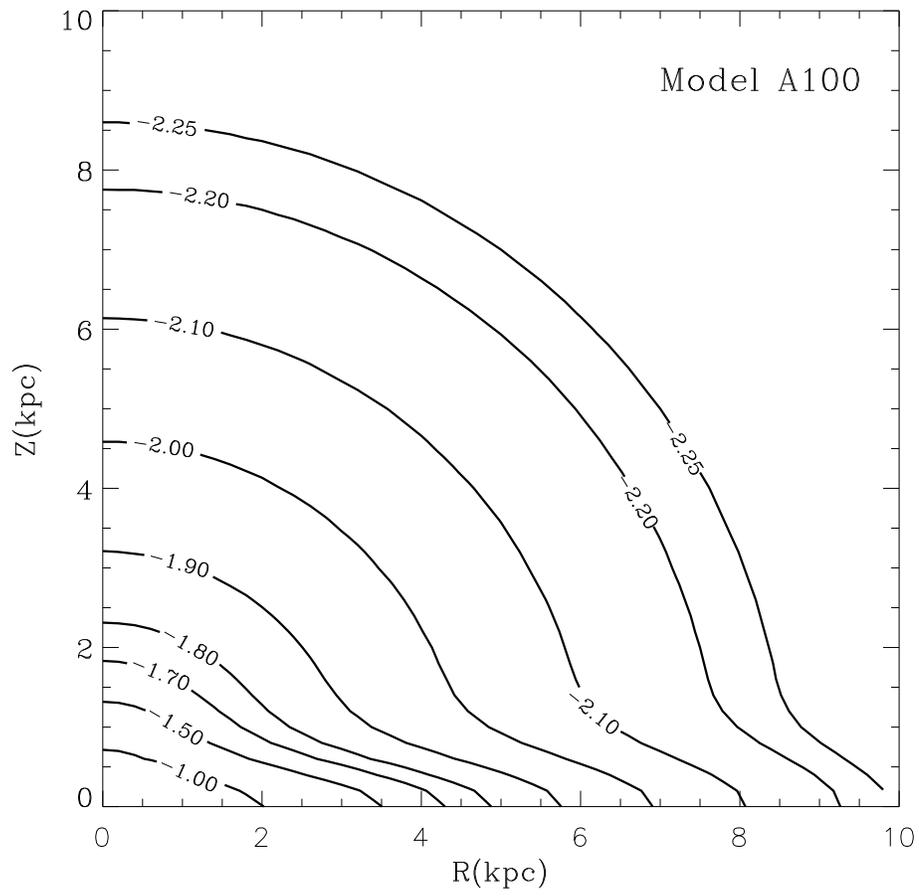
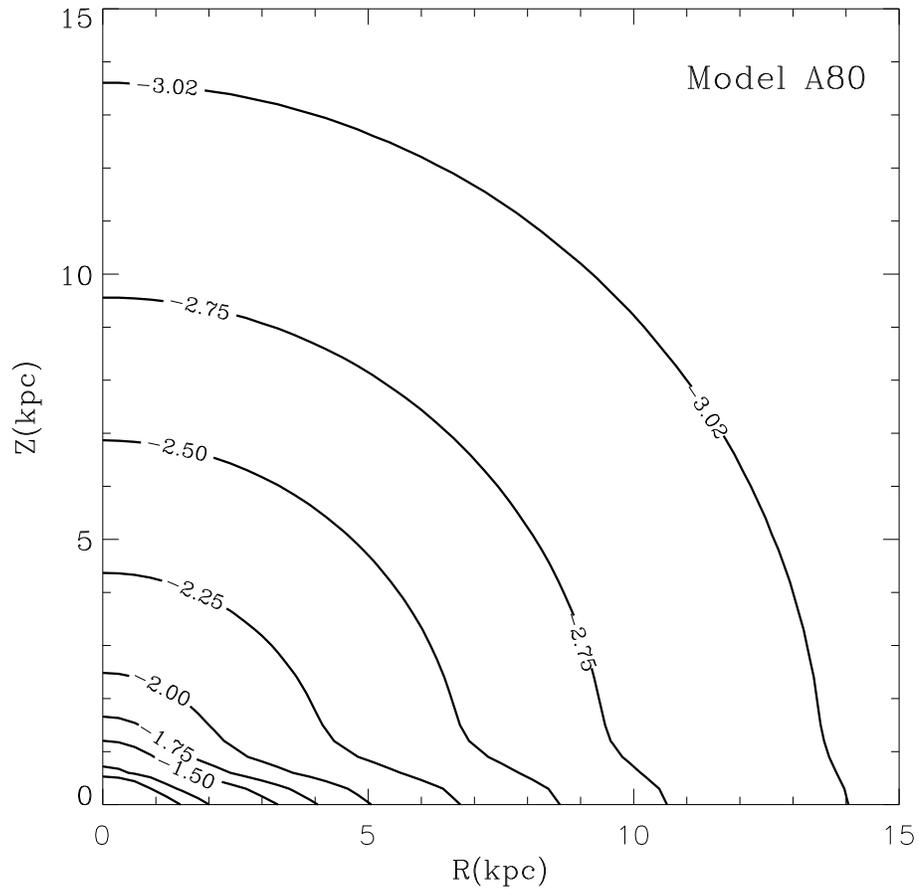
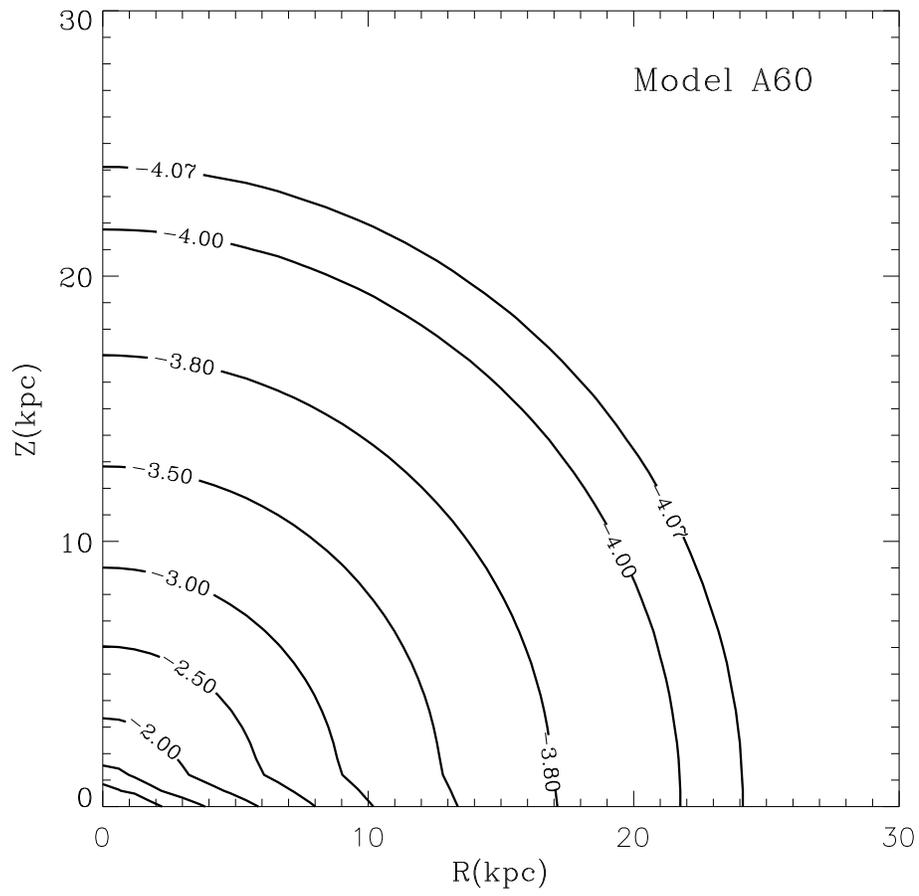
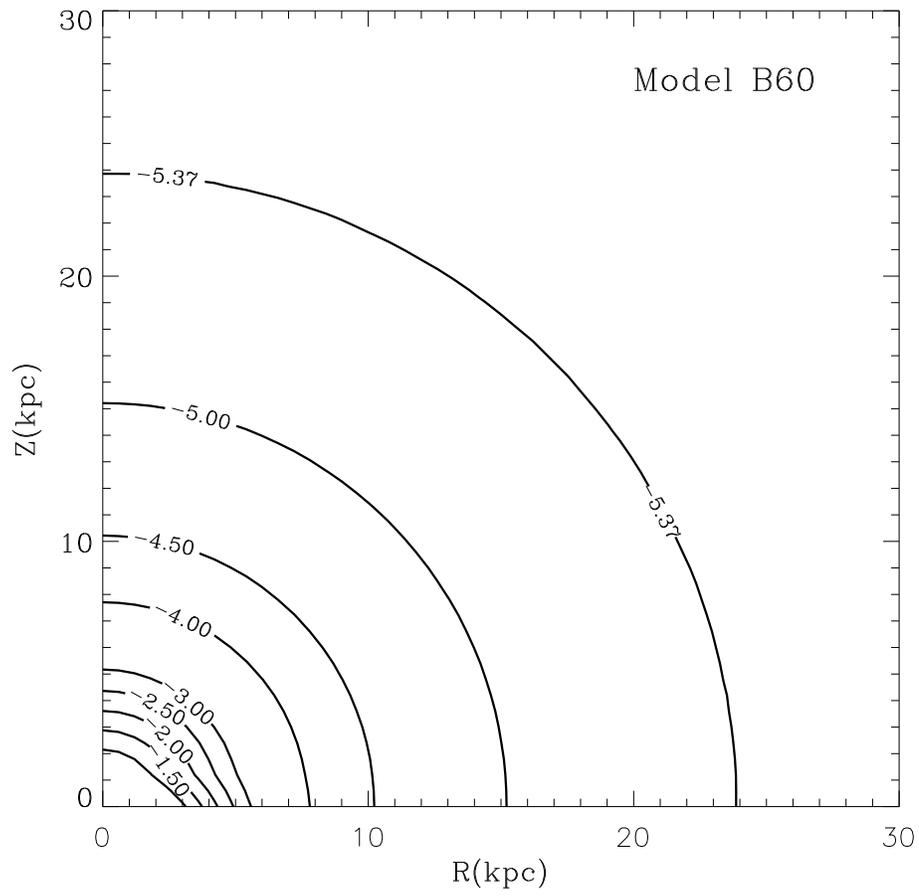
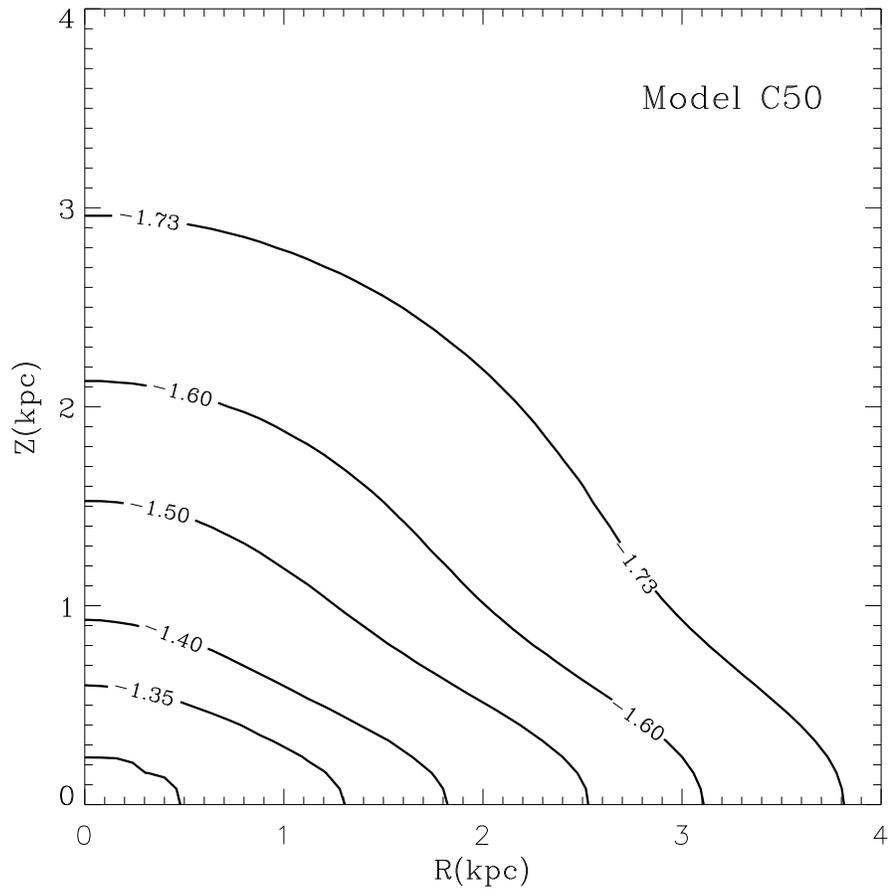
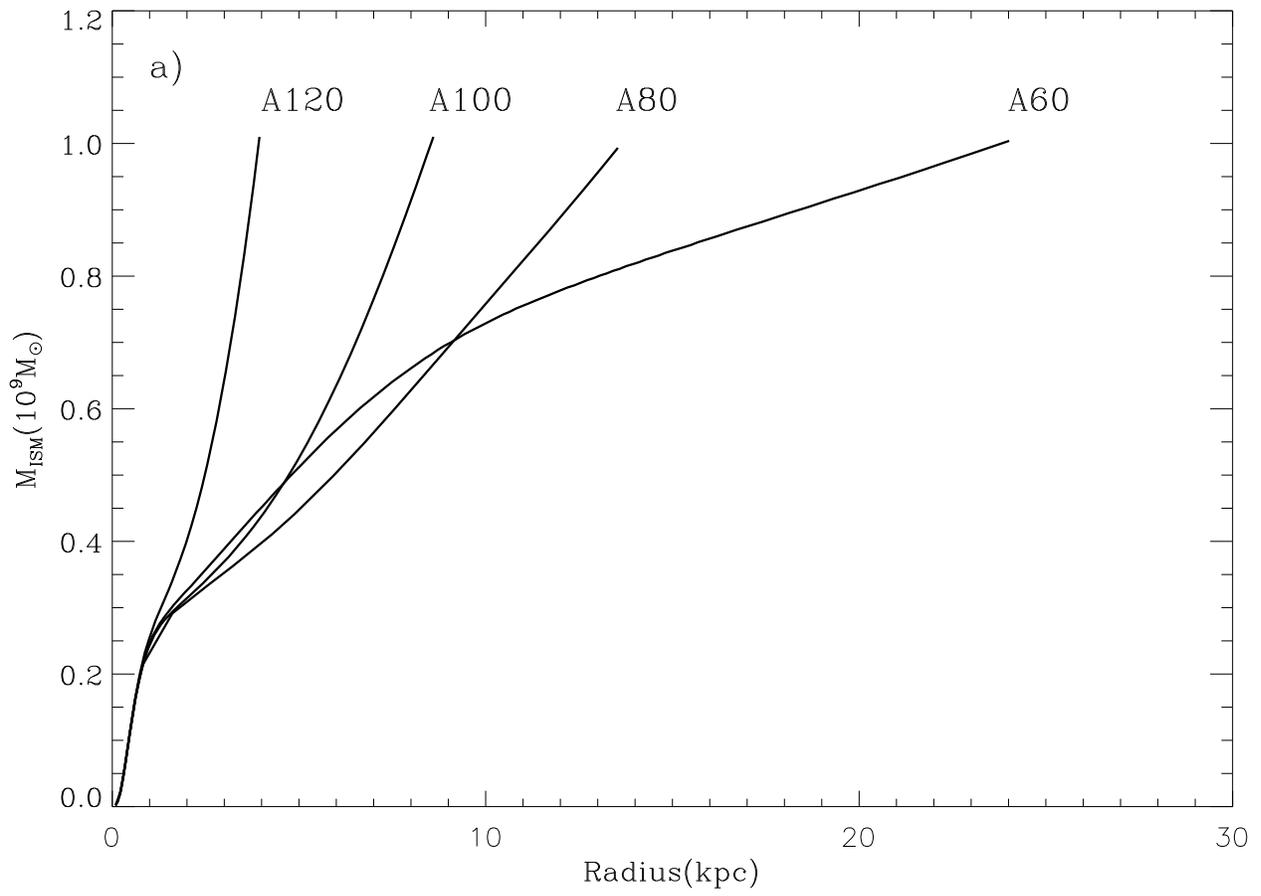
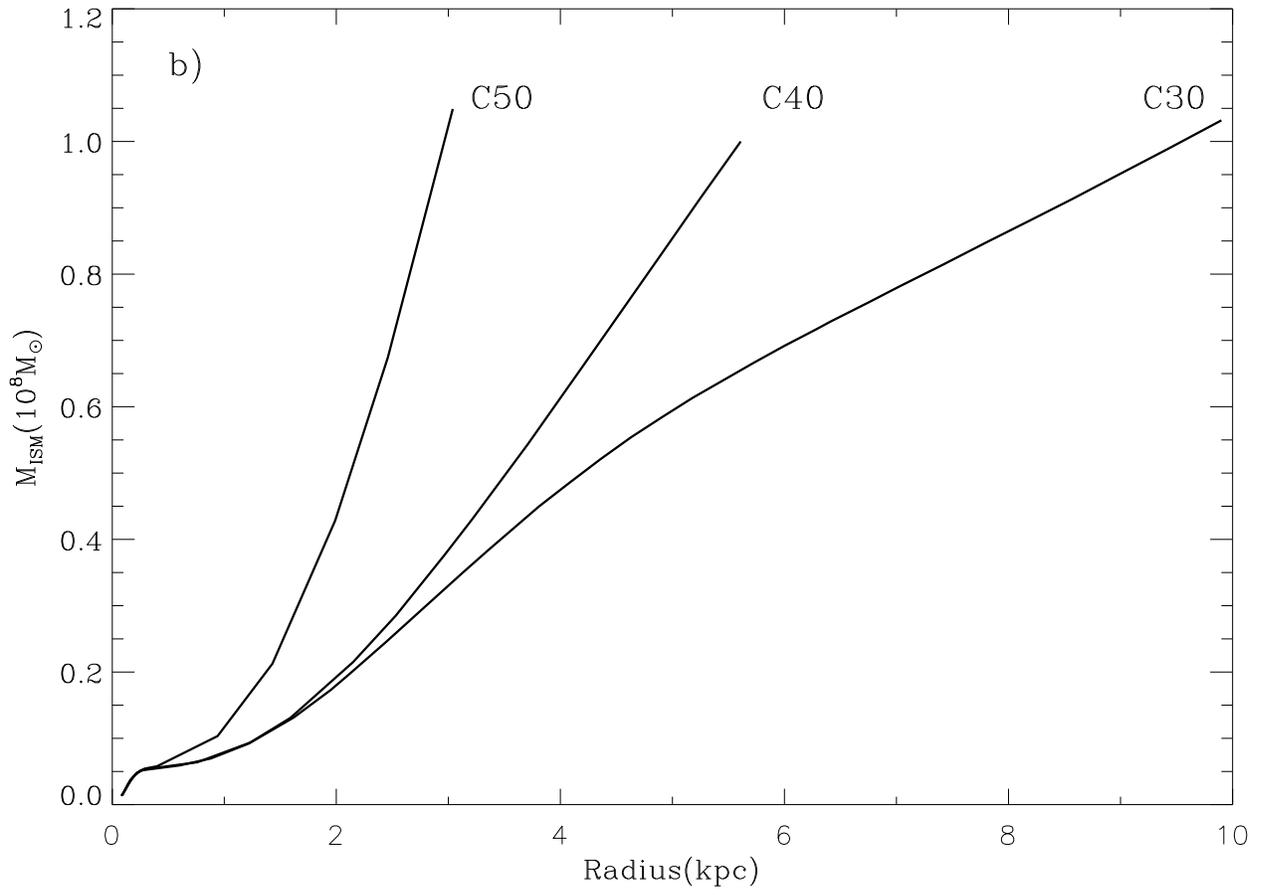
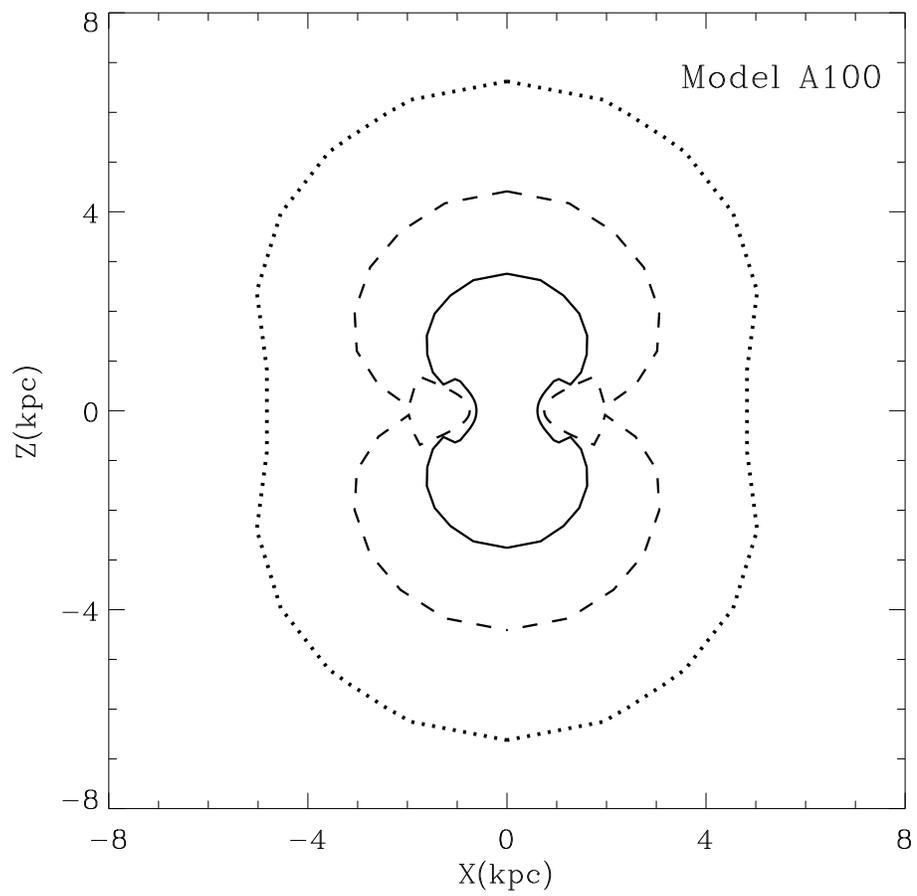
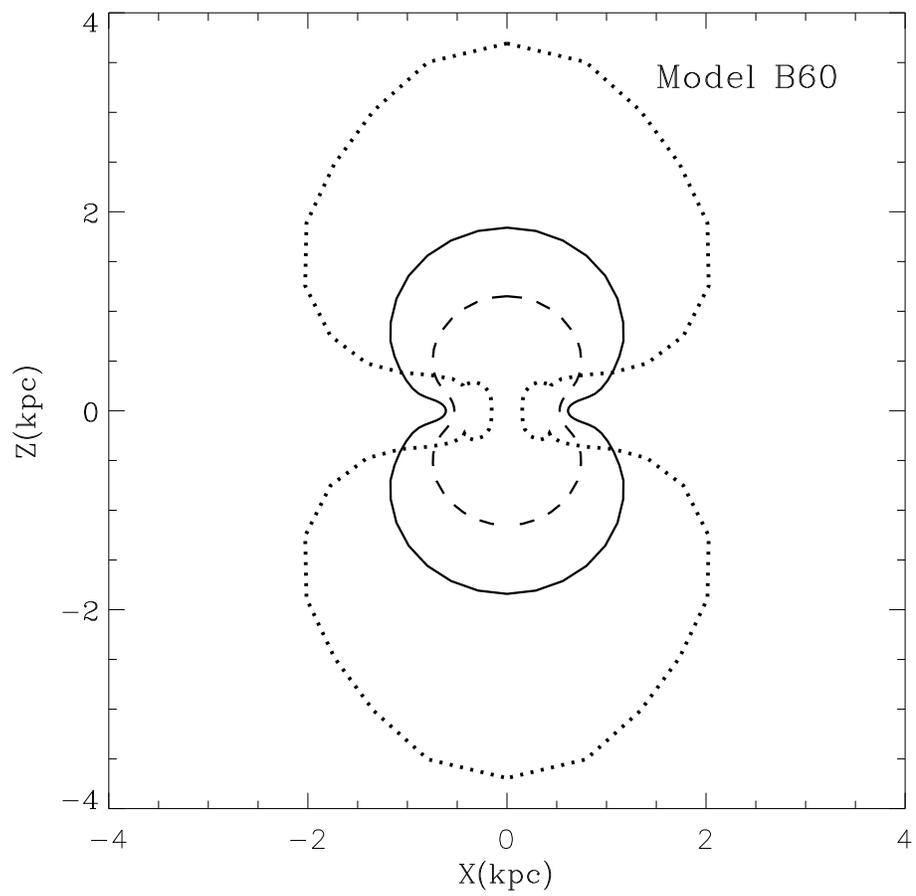
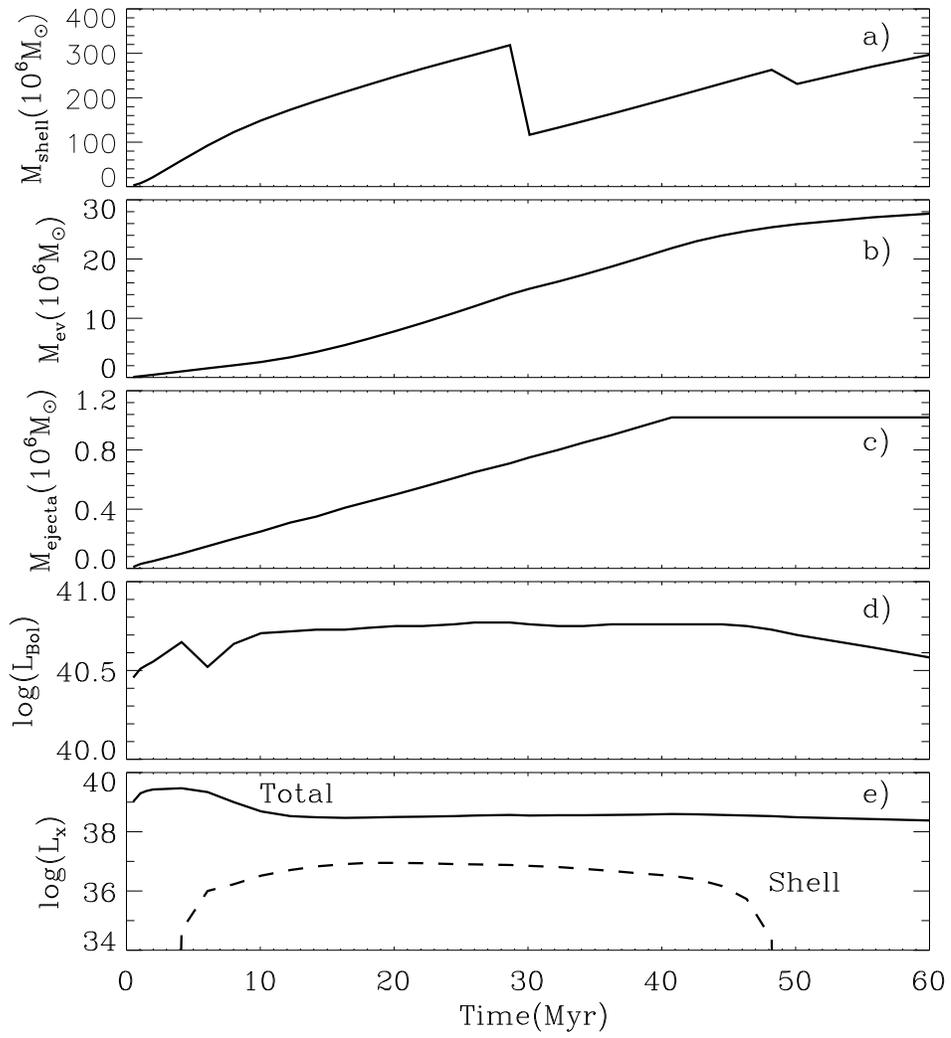
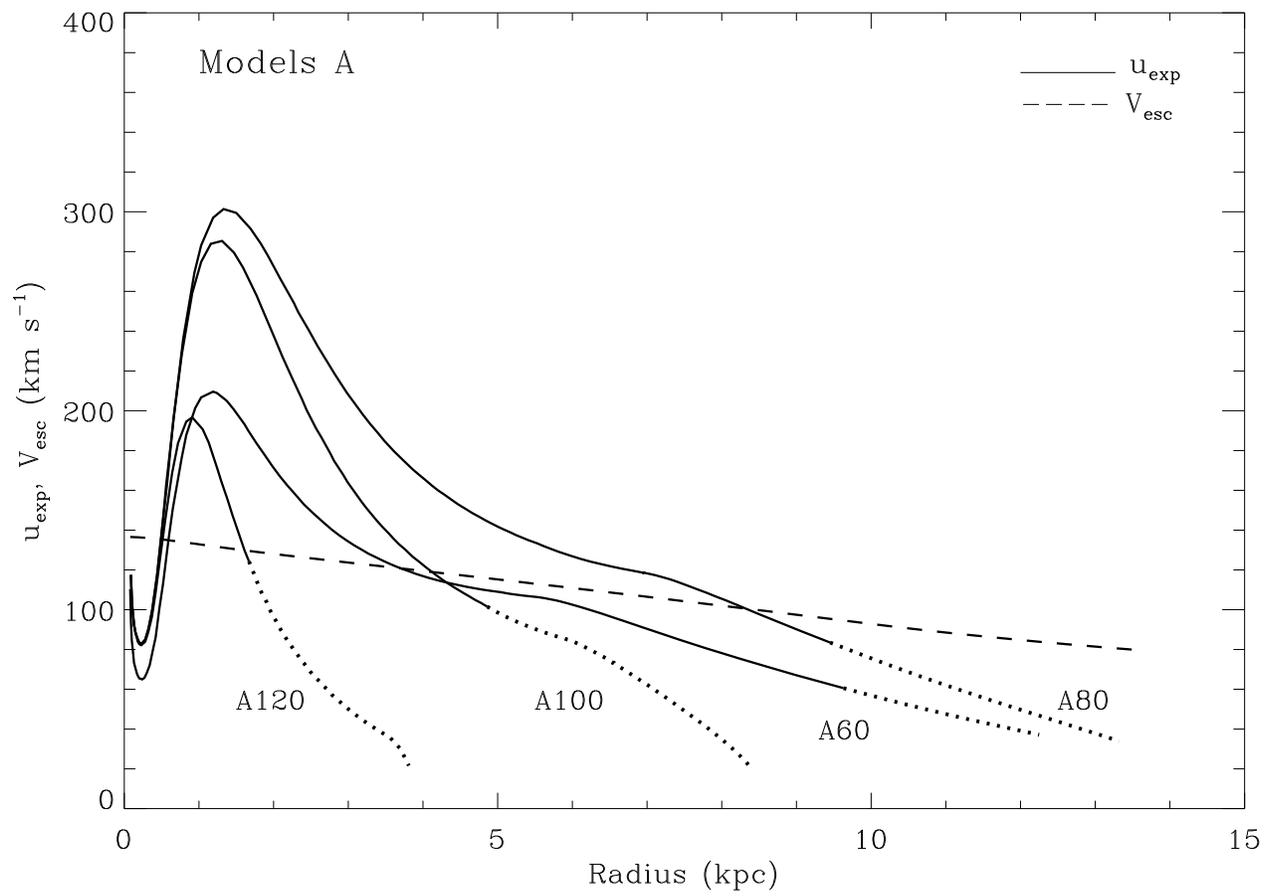
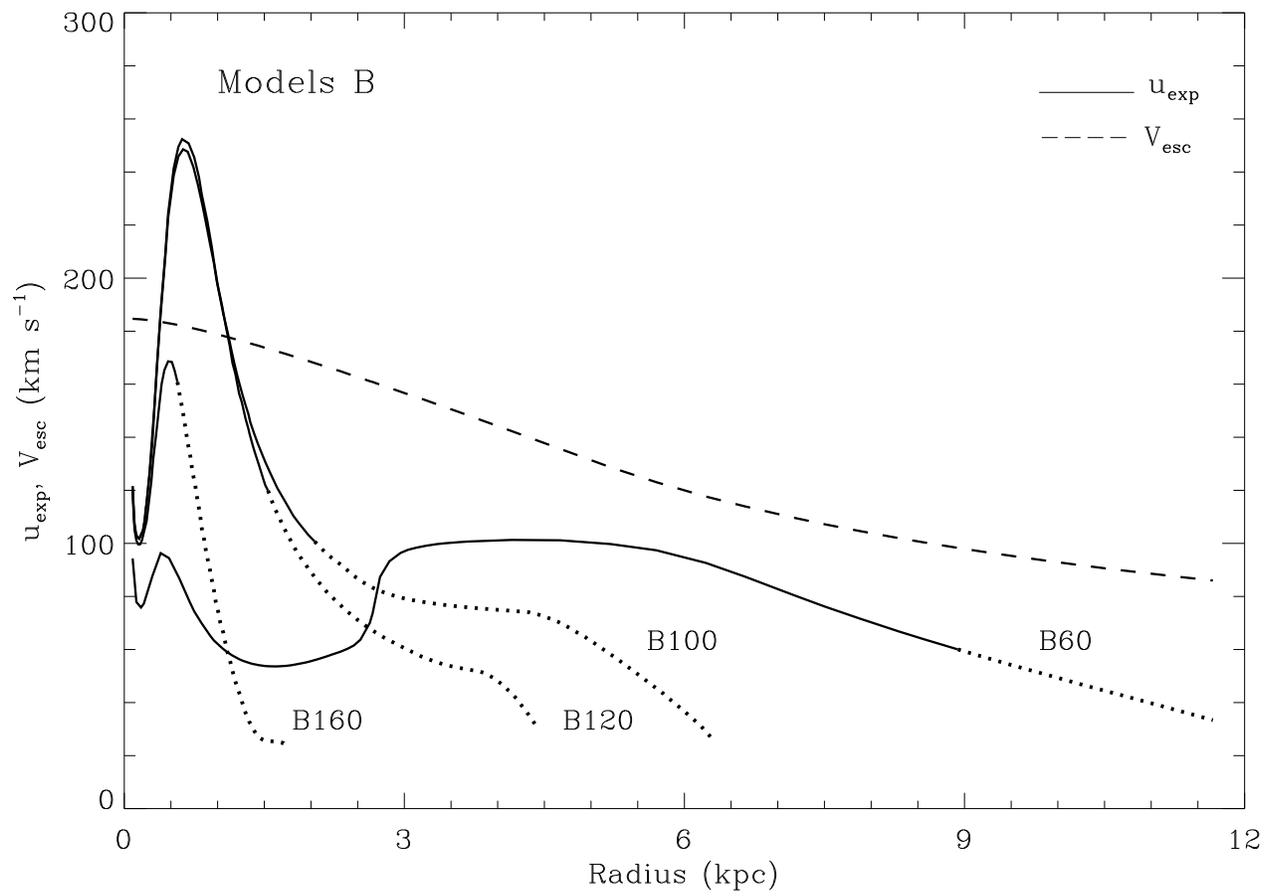
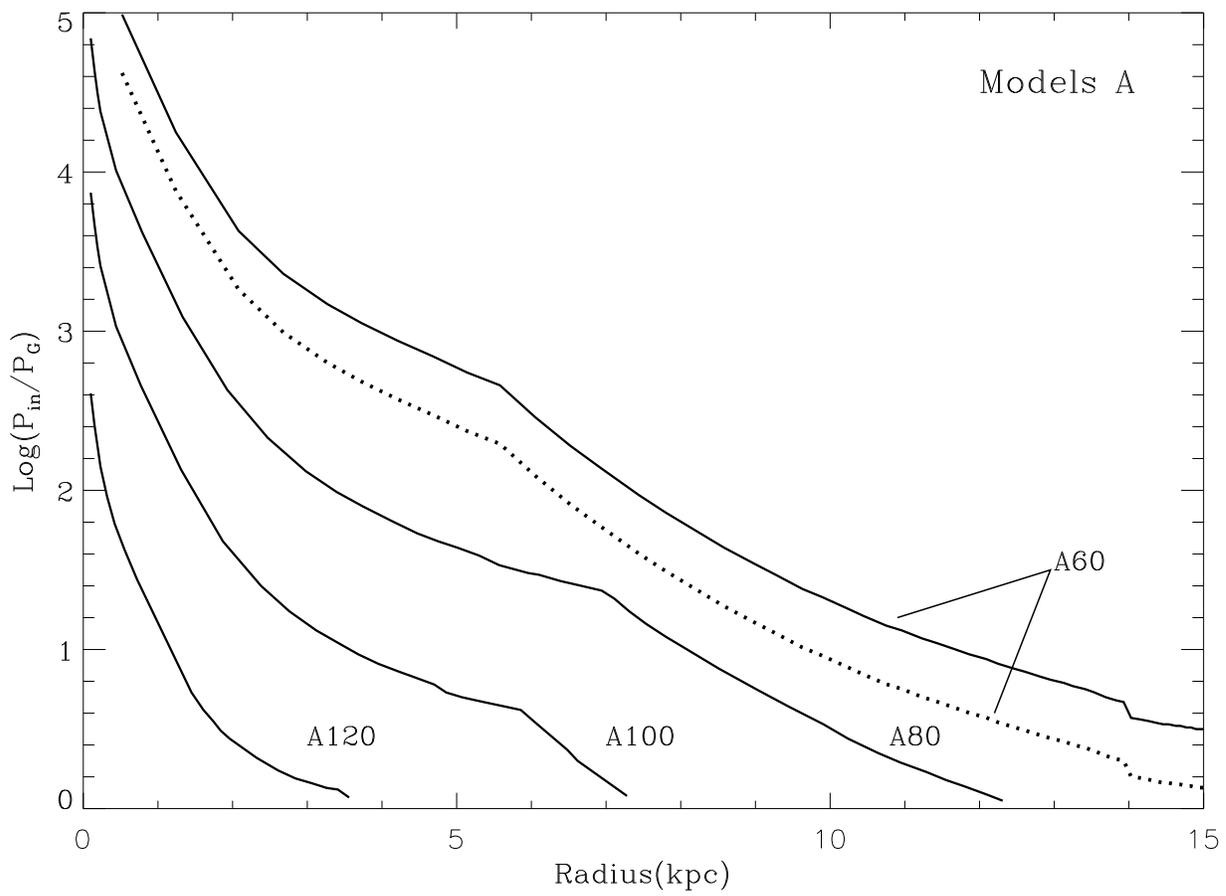
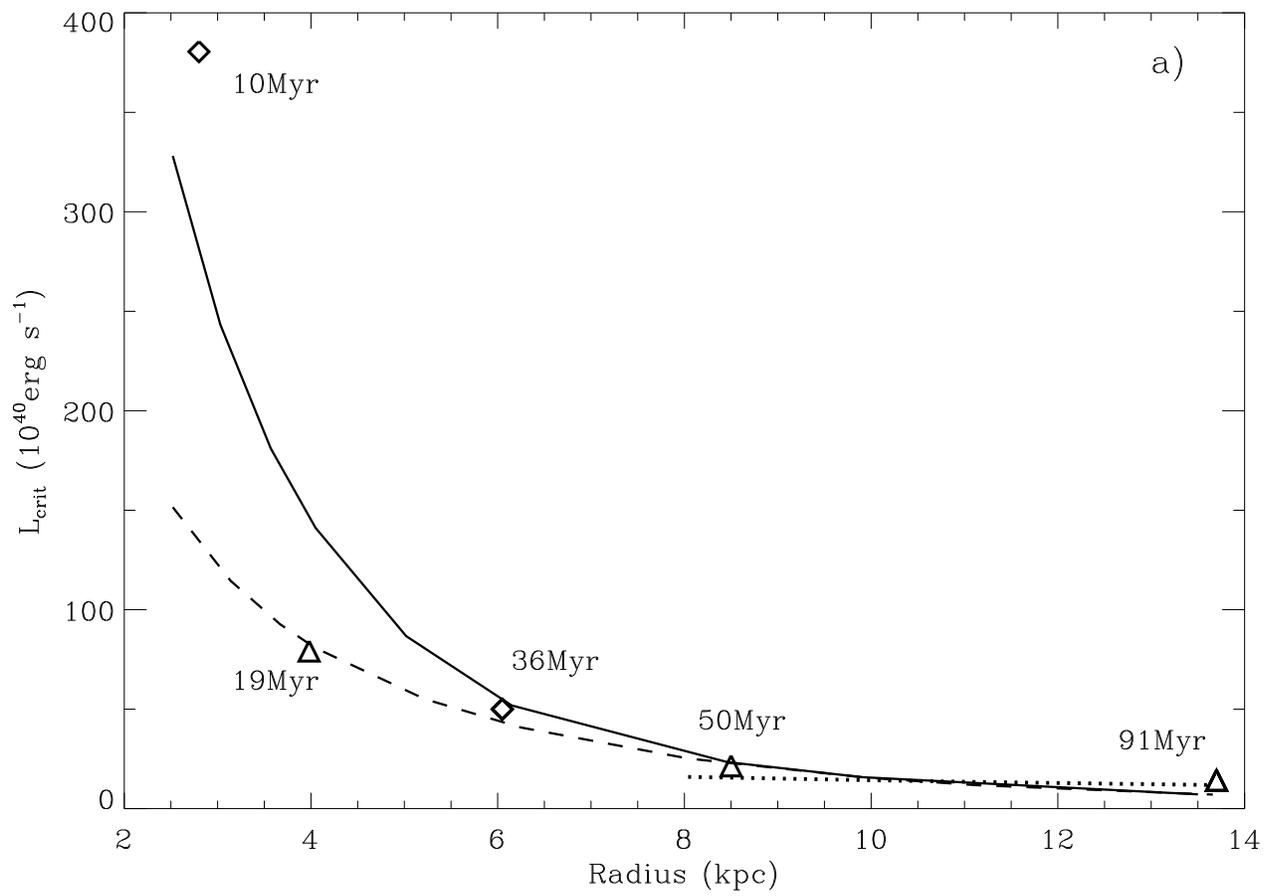
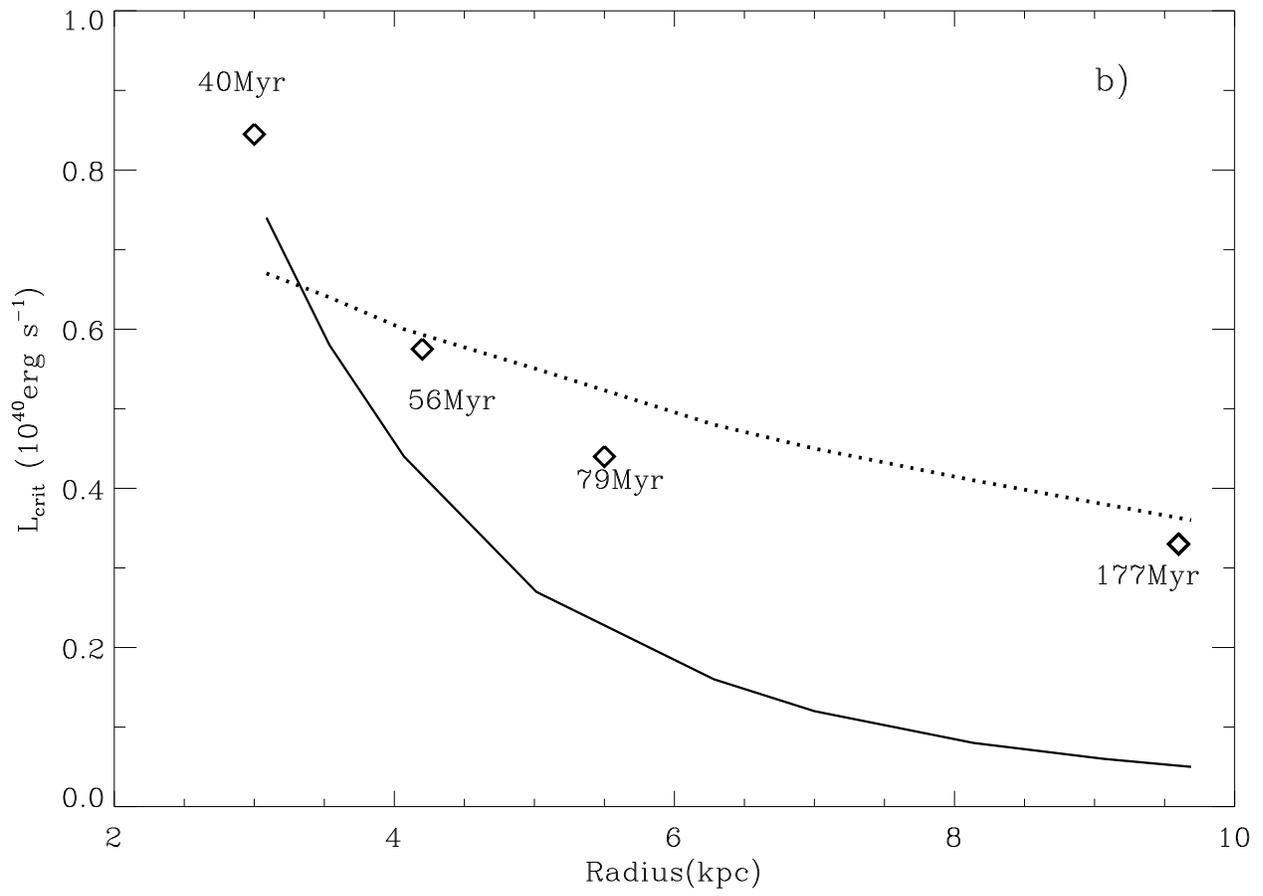
\end{document}